\def\year{2018}\relax
\documentclass[letterpaper]{article}
\usepackage{aaai18}
\usepackage{times}
\usepackage{verbatim}
\usepackage{graphicx}
\usepackage{subfigure}
\usepackage{multirow}
\usepackage{algorithmic}
\usepackage{longtable}
\usepackage{array}
\usepackage{stfloats}
\usepackage{balance}
\usepackage{multicol}
\usepackage{color}
\usepackage{epstopdf}
\usepackage{bm}
\usepackage{amsmath}
\usepackage{amssymb}
\usepackage{amsthm}
\usepackage{booktabs} %
\usepackage{epsfig}
\usepackage{enumitem}
\usepackage{cleveref}
\usepackage{xspace}
\usepackage{url}
\usepackage{float}
\usepackage{booktabs}
\usepackage{makecell}
\usepackage{graphicx}
\newcommand{\hide}[1]{} %
\newcommand{\vpara}[1]{\vspace{0.01in}\noindent\textbf{#1 }}
\newcommand{\para}[1]{\vspace{0.01in}\noindent\textbf{#1 }}
\newcommand{\secref}[1]{Section~\ref{#1}} %
\newcommand{\figref}[1]{Fig.~\ref{#1}} %
\newcommand{\tableref}[1]{Table~\ref{#1}}

\newcommand{\local}{local\xspace}
\newcommand{\migrant}{settled migrant\xspace}
\newcommand{\newcomer}{new migrant\xspace}
\newcommand{\locals}{locals\xspace}
\newcommand{\migrants}{settled migrants\xspace}
\newcommand{\newcomers}{new migrants\xspace}
\newcommand{\Locals}{Locals\xspace}
\newcommand{\Migrants}{Settled migrants\xspace}
\newcommand{\Newcomers}{New migrants\xspace}

\newcommand{\NEWCOMERS}{New Migrants\xspace}

\begin{document}

\title{Urban Dreams of Migrants: A Case Study of Migrant Integration in Shanghai}
\author{
Yang Yang$^{\dag}$, Chenhao Tan$^{\star}$, Zongtao Liu$^{\dag}$, Fei Wu$^{\dag}$, and Yueting Zhuang$^{\dag}$\\
$^{\dag}$College of Computer Science and Technology, Zhejiang University, China\\
$^{\star}$Department of Computer Science, University of Colorado Boulder, USA\\
yangya@zju.edu.cn, chenhao@chenhaot.com, \{tomstream, wufei, yzhuang\}@zju.edu.cn\\
}

\maketitle

\begin{abstract}

Unprecedented human mobility has driven the rapid urbanization 
around the world.
In China, the fraction of 
population dwelling in cities increased from 17.9\% to 52.6\% between 1978 and 2012.
Such large-scale migration poses %
challenges for policymakers and important questions for researchers.

To investigate the process of migrant integration, we employ a one-month complete dataset of telecommunication metadata in Shanghai with 54 million users and 698 million call logs. 
We find systematic differences between locals and migrants in their {\em mobile communication networks} and {\em geographical locations}.
For instance, migrants have more diverse contacts and move around the city with a larger radius than \locals after they settle down.
By distinguishing \newcomers (who recently moved to Shanghai) from \migrants (who have been in Shanghai for a while), we demonstrate the integration process of \newcomers in their first three weeks.
Moreover, we formulate classification problems to predict whether a person is a migrant. 
Our classifier is able to achieve an F1-score of 0.82 when distinguishing \migrants from \locals, but it remains challenging to identify \newcomers because of class imbalance. 
This classification setup holds promise for identifying \newcomers who will successfully integrate into \locals (\newcomers that misclassified as \locals).

\end{abstract}

\section{Introduction}
\label{sec:intro}

\begin{figure*}[t]
\centering
\subfigure[Overall average probability. \label{fig:classall_35_all_1_31}] {
\epsfig{file=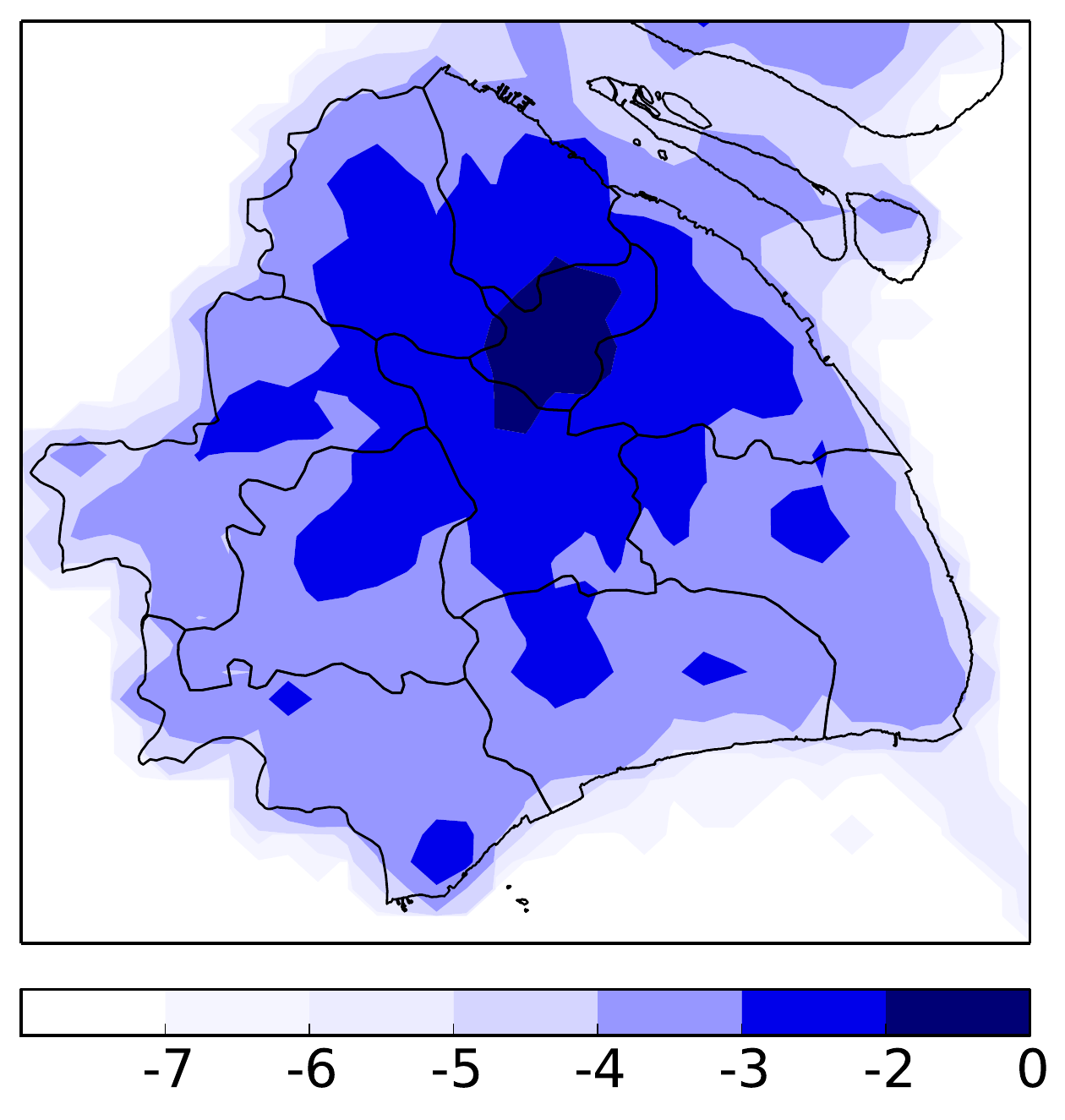, width=0.22\textwidth}
}
\hfill
\subfigure[\Locals. \label{fig:ratio_class1_all1_31}] {
\epsfig{file=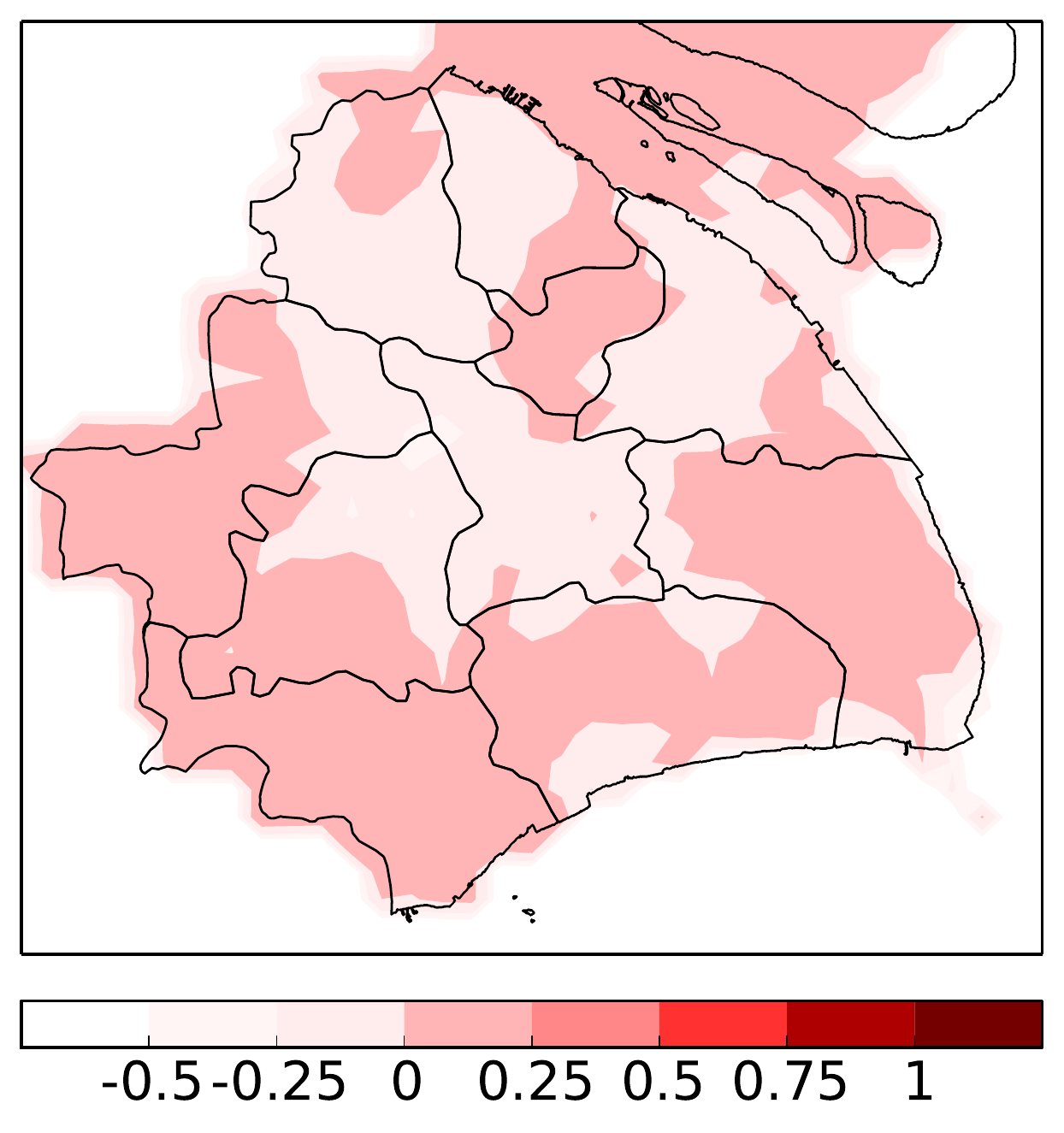, width=0.22\textwidth}
}
\hfill
\subfigure[Settled migrants.\label{fig:ratio_class2_all1_31}] {
\epsfig{file=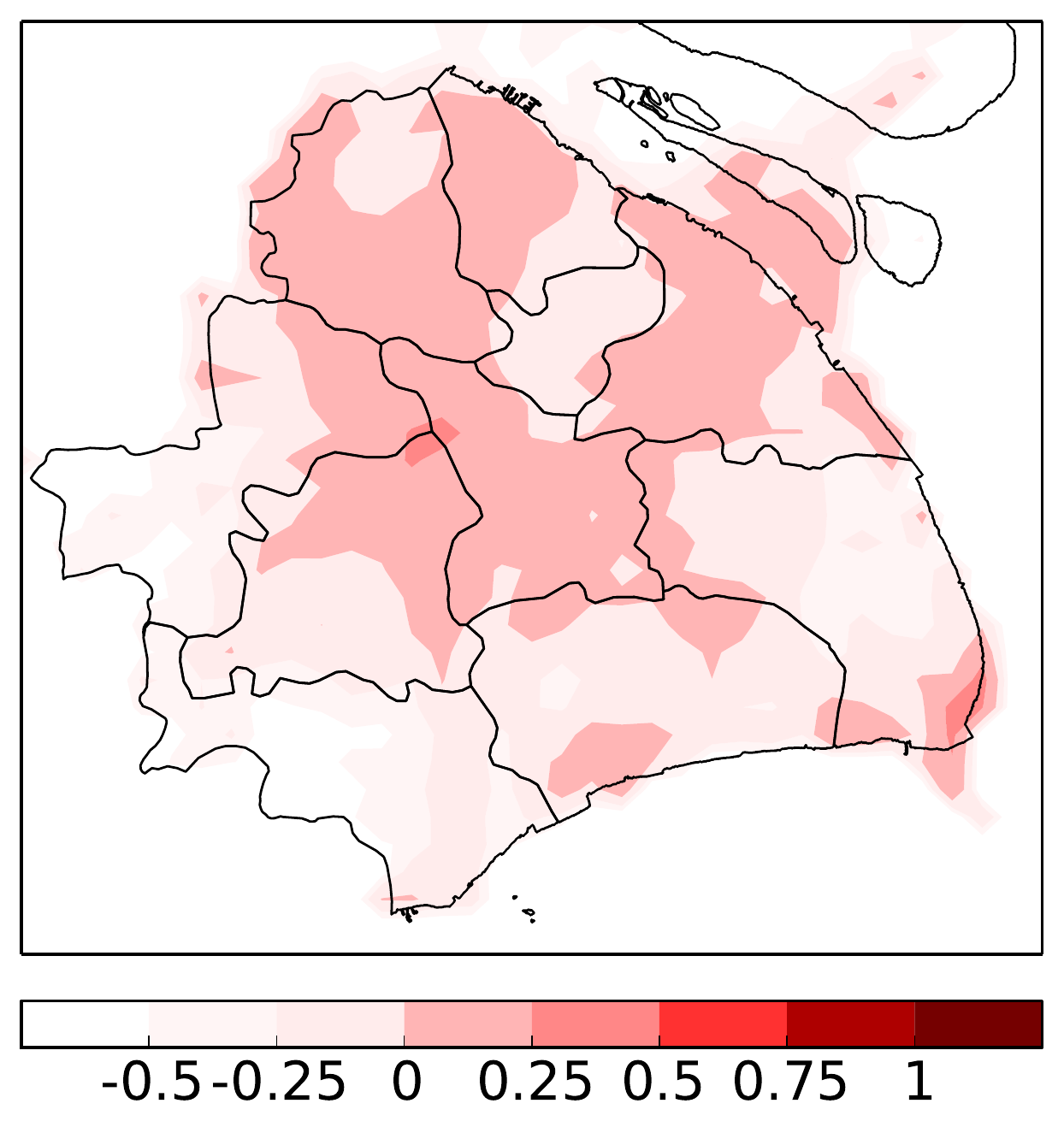, width=0.22\textwidth}
}
\hfill
\subfigure[New migrants. \label{fig:ratio_class3_all1_31}] {
\epsfig{file=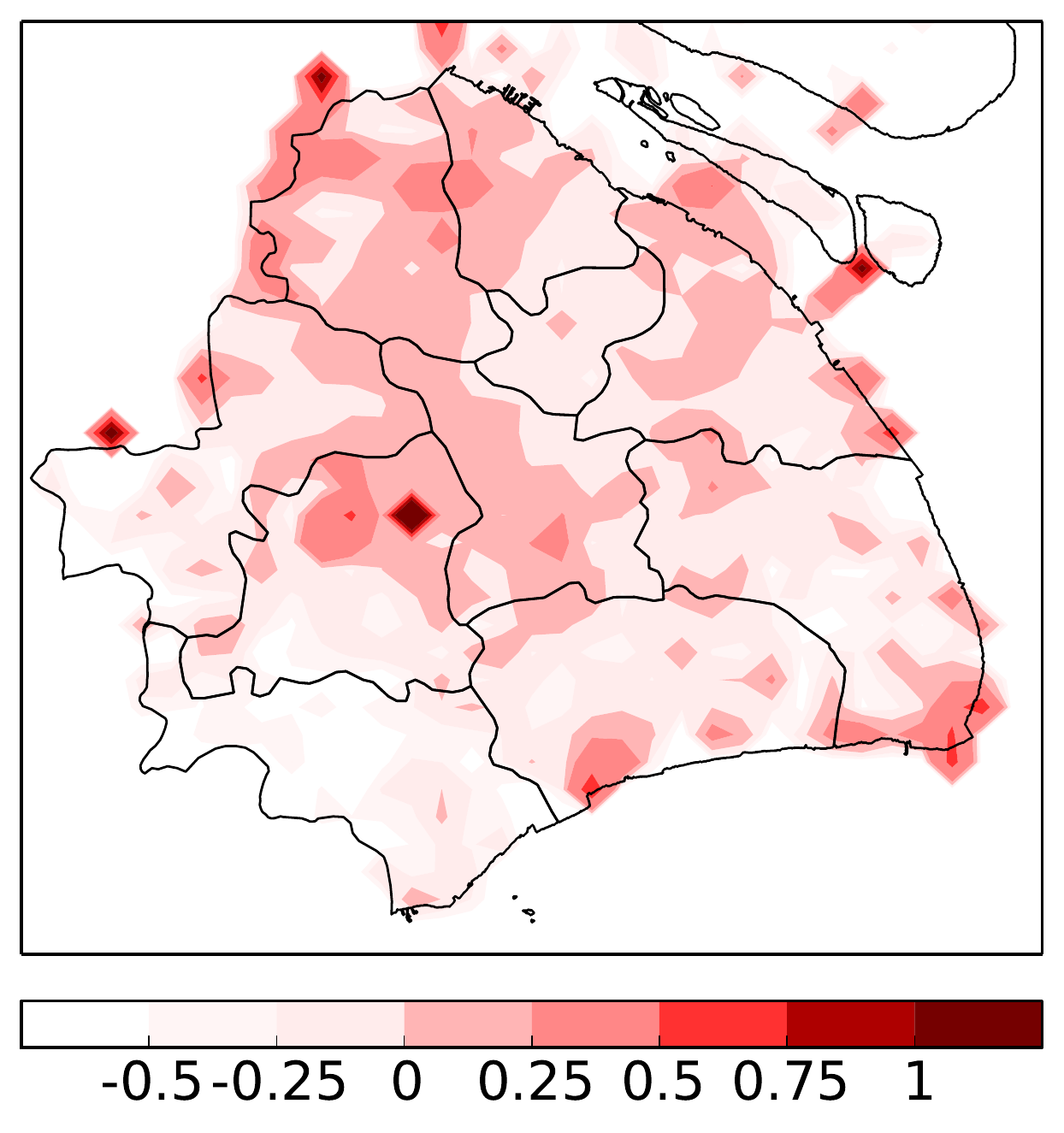, width=0.22\textwidth}
}
\caption{Geographical distributions of \locals, \migrants (who have been in a new city for a while), and \newcomers (who recently moved to a new city) in Shanghai. 
\small 
Each person is represented by the center of their active areas.
\figref{fig:classall_35_all_1_31} shows the log probability of all users in each region and this probability constitutes a comparison point for the other figures.
Each of the right three figures shows the log odds ratio of each group compared to the overall average in \figref{fig:classall_35_all_1_31}, i.e., $\log P_{\operatorname{group}} - \log \overline{P}$, where $\overline{P}$ is the overall average in \figref{fig:classall_35_all_1_31} and $P_{\operatorname{group}}, \operatorname{group} \in \{\operatorname{\locals}, \operatorname{settled~migrants}, \operatorname{new~migrants}\}$ is the probability to fall in a region within that particular group of people.
Intuitively, a red region in the right three figures suggests that this group of people are disproportionally frequent in that region.
\Migrants tend to be in the central part of the city, while \locals are in the periphery.
\Newcomers are similar to \migrants, but have a few dark areas on the periphery.
The darkest point in \figref{fig:ratio_class3_all1_31} correspond to Songjiang University Town, a hub of universities. 
\normalsize
}
\label{fig:geo}
\end{figure*}

More than half of the world's population are now living in urban areas~\cite{nations2014world}.
This rapid urbanization process involves a continuous flow of migrants into the cities.
For example, the number of migrants now live in China is 236 million, 17\% of the country's entire population~\cite{Lin:2013}. 
These migrants play an important role in a city's rapid development by strengthening its political and economic status and bringing diverse cultures to the city \cite{lee2015world}. 
However, great challenges arise because of the fast rate of migration.
Policymakers need to address a multitude of issues regarding migrants in modern cities, including environment, land, labor, segregation, and social inequality~\cite{Bai:NatureNews:2014,lee2015world,razavi2010underpaid,Goodburn:InternationalJournalOfEducationalDevelopment:2009}.
It is thus an important research question to understand how migrants integrate into a city. 

In this work, 
we are interested in two central components of migrant integration: the locations where a migrant lives and moves around, and people that a migrant interacts with and befriend.
First, because cities are divided into neighborhoods with varying characteristics, there may exist systematic differences between \locals and migrants in where they live.
For example, \figref{fig:geo} shows the geographical distributions of \locals and migrants compared to the overall average in Shanghai.
Somewhat surprisingly, \locals are more active in the periphery of the city,
whereas migrants relatively concentrate in the center of Shanghai. 
This observation echoes previous findings that existing residents flee from central cities, known as ``white flight'' \cite{frey1979central}.
It yet remains an open question how migrants' active areas evolve as they integrate into the city.

Another important aspect of 
migrant integration is how migrants establish their personal connections.
As humans are social animals, whether a migrant can successfully develop a personal network is crucial in her integration process \cite{gurak1992migration}.
In particular, \citeauthor{yue2013role} show that migrant-resident ties are significantly associated with migrant integration~\cite{yue2013role}.
However, it remains unclear how a migrant makes initial friends and then gradually build a personal network in a new city.
It is also unknown
what characteristics differentiate the social networks of migrants from those of \locals.

In order to investigate the above two aspects, we conduct a case study of Shanghai, one of the biggest cities in China, and present the first large-scale quantitative exploration 
of migrant integration.
We employ a {\em one-month complete} dataset of telecommunication metadata from China Telecom,\footnote{China Telecom Corporation is a Chinese state-owned telecommunication company and the third largest mobile service providers in China.} which contains 698 million call logs between  
54 million mobile users in Shanghai.
To identify a comparison point of migrants, we define {\em locals} as the persons that were born in Shanghai, the counterpart of migrants.
As migrants may undergo different stages in their integration,
we further differentiate migrants that have been in a new city for a while, {\em \migrants}, from migrants that recently moved to a new city, {\em \newcomers} (\secref{sec:setup}).

First, we explore how \locals, \migrants, and \newcomers differ in their mobile communication networks and geographical locations in \secref{sec:analysis}.
We find interesting differences between these three groups.
For instance, in terms of communication networks, a substantial fraction of \newcomers' contacts are fellow townsmen, people who were born in the same province.
This pattern suggests that townsmen are essential for \newcomers to build their initial personal networks in a new city.
Surprisingly, \migrants have an even higher fraction of townsman contacts, indicating that they may have grown their townsman network as they stay longer in Shanghai.
In terms of locations, in addition to the differences in \figref{fig:geo},
we find that \migrants tend to have a larger radius.

\begin{table}[!t]
 \centering 
  \small
 \begin{tabular}{lp{0.3\textwidth}}
 \toprule
 Feature &  Description \\
 \midrule
\multicolumn{2}{c}{\textbf{Demographics of user $v$'s friends in $G_t$}
} \\
 similar-age & The fraction of $v$'s friends that are at similar ages with $v$ ($\pm 10$ years). \\
 same-sex & The fraction of $v$'s friends having the same sex with $v$. \\
 local & The fraction of $v$'s friends who were born in Shanghai. \\
 townsmen & The fraction of $v$'s friends that were born in the same province with $v$ but not in Shanghai.
 \\
 \midrule
 \multicolumn{2}{c}{\textbf{Ego-network characteristics of user $v$ in $G_t$}} \\
 degree & The number of $v$'s unique contacts. \\ 
 weighted degree & The number of calls $v$ makes. \\
 neighbor degree & The average degree of $v$'s contacts. \\
 CC  & Local clustering coefficient of $v$, i.e., $\frac{|\{(s, t)| (s, t) \in E_t\}|}{d_v(d_{v}-1)}$, where $s$ and $t$ are $v$'s friends, and $d_{v}$ is $v$'s degree. \\
 \midrule 
 \multicolumn{2}{c}{\textbf{Call behavior in $G_t$}} \\
 call duration & $v$'s average call duration. \\
 duration variance & variance of $v$'s call duration.\\
 province diversity & Entropy of the birth provinces distribution 
 among $v$'s contacts, 
 $-\sum_i p_i \log_2 p_i$, where $p_i$ is the probability that $v$'s contact was born in province $i$.\\
 reciprocal call & The probability that $v$'s contacts also call $v$ in week $t$.
 \\
 \midrule 
 \multicolumn{2}{c}{\textbf{Geographical features of $v$ at week $t$}} \\
 center & The latitude and longitude of a user $v$'s center of mass $l_{\operatorname{CM}}$,
 $l_{\operatorname{CM}}=\frac{1}{|L_v^t|}\sum_{l \in L_v^t}l$.
 \\
max radius & The maximal distance of $v$ from her center of mass, i.e., $\max_{l \in L_v^t} |l - l_{\operatorname{CM}}|$. \\
 average radius & The average distance of $v$ from her center of mass, i.e., $\frac{1}{|L_v^t|}\sum_{l \in L_v^t} |l - l_{\operatorname{CM}}|$.\\
 moving distance & The total distance that $v$ moves, i.e., $\sum_i |l_i - l_{i-1}|$. \\
 average distance & The average distance that $v$ moves, i.e., $\frac{1}{|L_v^t|}\sum_i |l_i - l_{i-1}|$. \\
 \bottomrule 
 \end{tabular}
 %
  \normalsize
   \caption{\label{tb:feature} 
 List of features in this paper. 
 \small 
 We view all directed edges as undirected except in measuring reciprocal calls.
 For demographics related features, we only include users for whom we have the corresponding information.
 \normalsize }
 \end{table}

Second, we use the calling logs over different time periods
 to give a brief dynamic view of the integration process in \secref{sec:integration}.
Despite the short time span, we observe that \newcomers 
become increasingly similar to \migrants in most characteristics, while features of \migrants and \locals tend to be stable over time.
This contrast suggests that the features that we employ can indeed reflect the integration process to some extent.
Meanwhile, we observe that the integration slows down in the final week.
One possible explanation is that not all \newcomers eventually become \migrants and the slow integration is due to the ones that encounter 
difficulty fitting into the city.
This hypothesis is worth further investigation.

Finally, we formulate prediction tasks to distinguish migrants from \locals in \secref{sec:prediction}. 
Using the features that we propose, we are able to build a classifier that significantly outperforms the baselines and achieve an F1-score of $0.82$ on predicting \migrants, indicating that it is not a difficult prediction task to separate \migrants from \locals.
We also observe that if we apply this classifier to \newcomers, an increasing fraction of \newcomers is classified as \locals over time.
However, it remains challenging to identify \newcomers because
the number of \newcomers is very small compared to \migrants and \locals.

Our work is a first step towards understanding migrant integration and informing urban policymakers.
We 
provide an overview of related work on this issue in \secref{sec:related} and offer some concluding discussions in \secref{sec:conclusion}.

\section{Experimental Setup}
\label{sec:setup}

In this section, we introduce our dataset and the framework that we use to study mobile communication networks and geographical information of \locals and migrants.

\subsection{Dataset}
\label{sec:data}
Our dataset contains \textit{complete} telecommunication records between mobile users using China Telecom in Shanghai, spanning a month from September 3rd, 2016, to September 30th, 2016 (four weeks). 
The data is provided by China Telecom,
the third largest mobile service provider in China. 
Our dataset consists of 
about 54 million 
users and 
698 million 
call logs between them.
A call log was recorded as long as it was made in Shanghai and either the caller or the callee was a user of China Telecom (some of the 54 million users use other mobile services).
Each call log contains the caller's number, the callee's number,
the starting time, and the ending time.
Since personal identification is required to obtain a mobile number,
we are able to retrieve personal attributes, including age, sex, and birthplace, for users of China Telecom that opened their accounts in Shanghai.\footnote{We obtain a person's birthplace based on the personal identity card number.}
Moreover, we can differentiate local numbers in Shanghai from numbers in other regions and getting a local number is a first step in the migrant integration process due to long-distance costs.
In addition, we have 
the GPS location of the corresponding telecommunications tower used 
during the call for users of China Telecom, which roughly approximates the locations of them.
Our dataset was anonymized by China Telecom for privacy concerns.
Throughout the paper, we report only average statistics without revealing any identifiable information of individuals.

\subsection{Framework}
\label{sec:notation}

We categorize users in our dataset into three groups based on their birthplaces and this categorization constitutes the basis for our computational framework.
We refer to people that were born in Shanghai as \textit{\locals}.
The rest people who were not born in Shanghai are migrants.
To assess different stages of migrant integration,
we separate migrants that have no call logs in the first week ({\em \newcomers}), from migrants that have at least one call log in the first week ({\em \migrants}). 
We further require each \local and \migrant to have call logs at every week, and each \newcomer to have call logs at each of the last three weeks,
to make sure that 
these users lived in Shanghai during our four-week span.
We filtered around $15,000$ users that have abnormally high degrees, who likely corresponded to fraudsters, delivery persons, or customer services according to a user type list provided by China Telecom.
In the end, we have {\em 1.7M \locals, 1.0M \migrants, and 22K \newcomers}.

One concern is that \newcomers in our dataset are simply temporal visitors to Shanghai. 
However, obtaining a phone number is nontrivial and requires personal identification in China, so it is uncommon for a temporary visitor to obtain a local number.

\begin{figure*}[t]
\centering
\subfigure[Demographics of friends.\label{fig:feature_attribute}] {
\epsfig{file=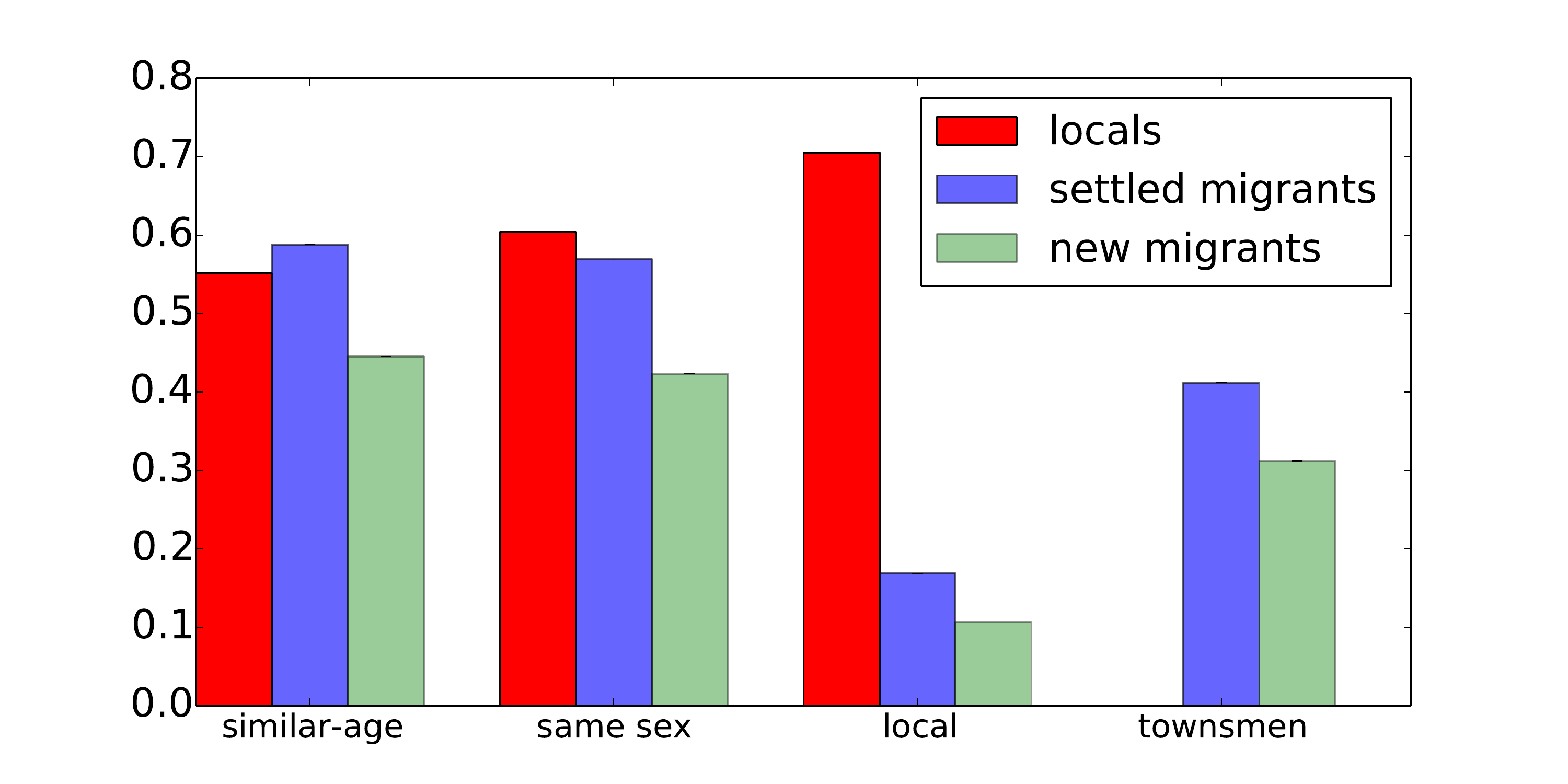, width=0.4\textwidth}
}
\subfigure[Ego-network characteristics.\label{fig:feature_network}] {
\epsfig{file=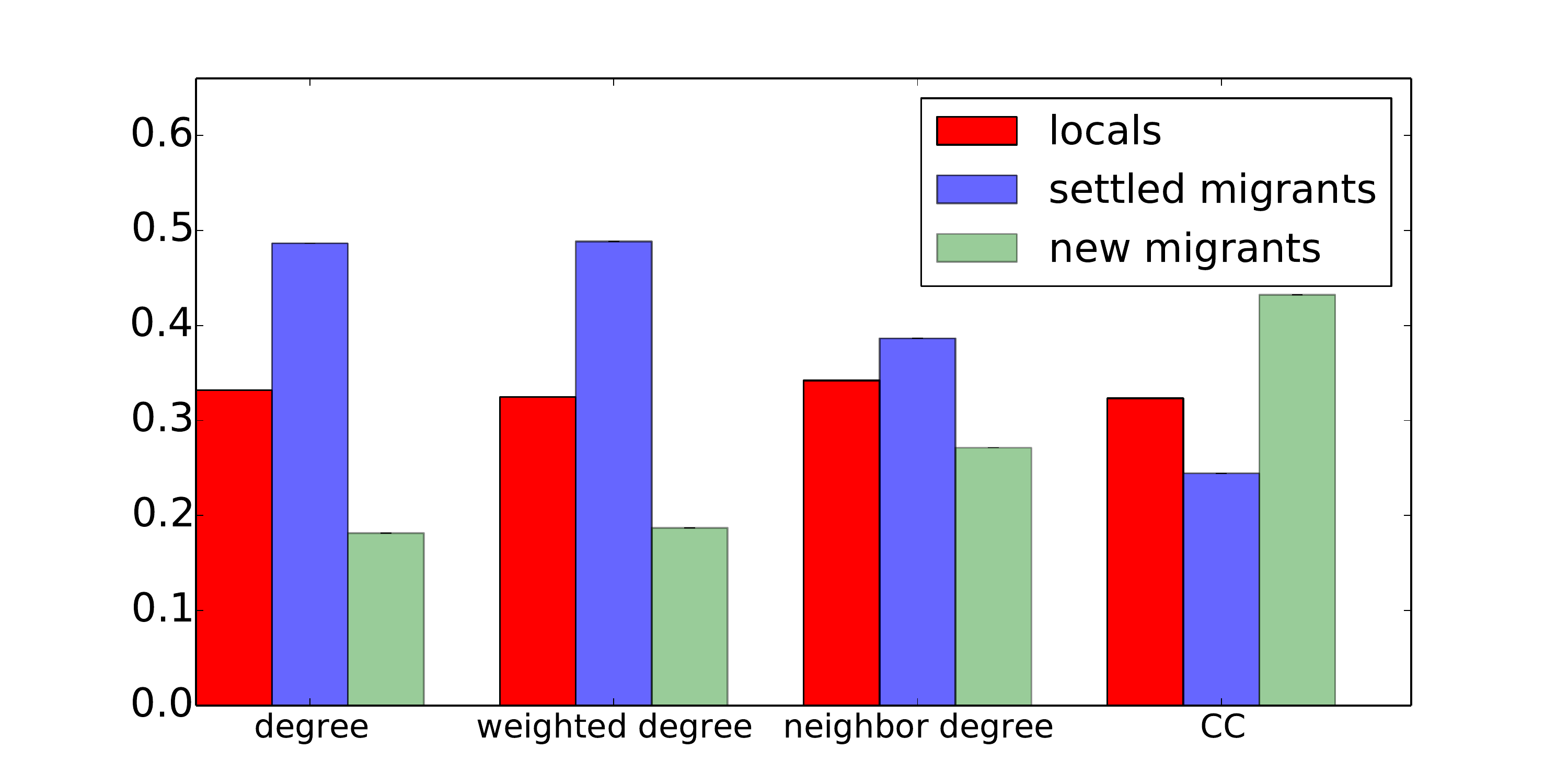, width=0.4\textwidth}
}\\
\vspace{-0.15in}
\subfigure[Call behavior.\label{fig:feature_call}] {
\epsfig{file=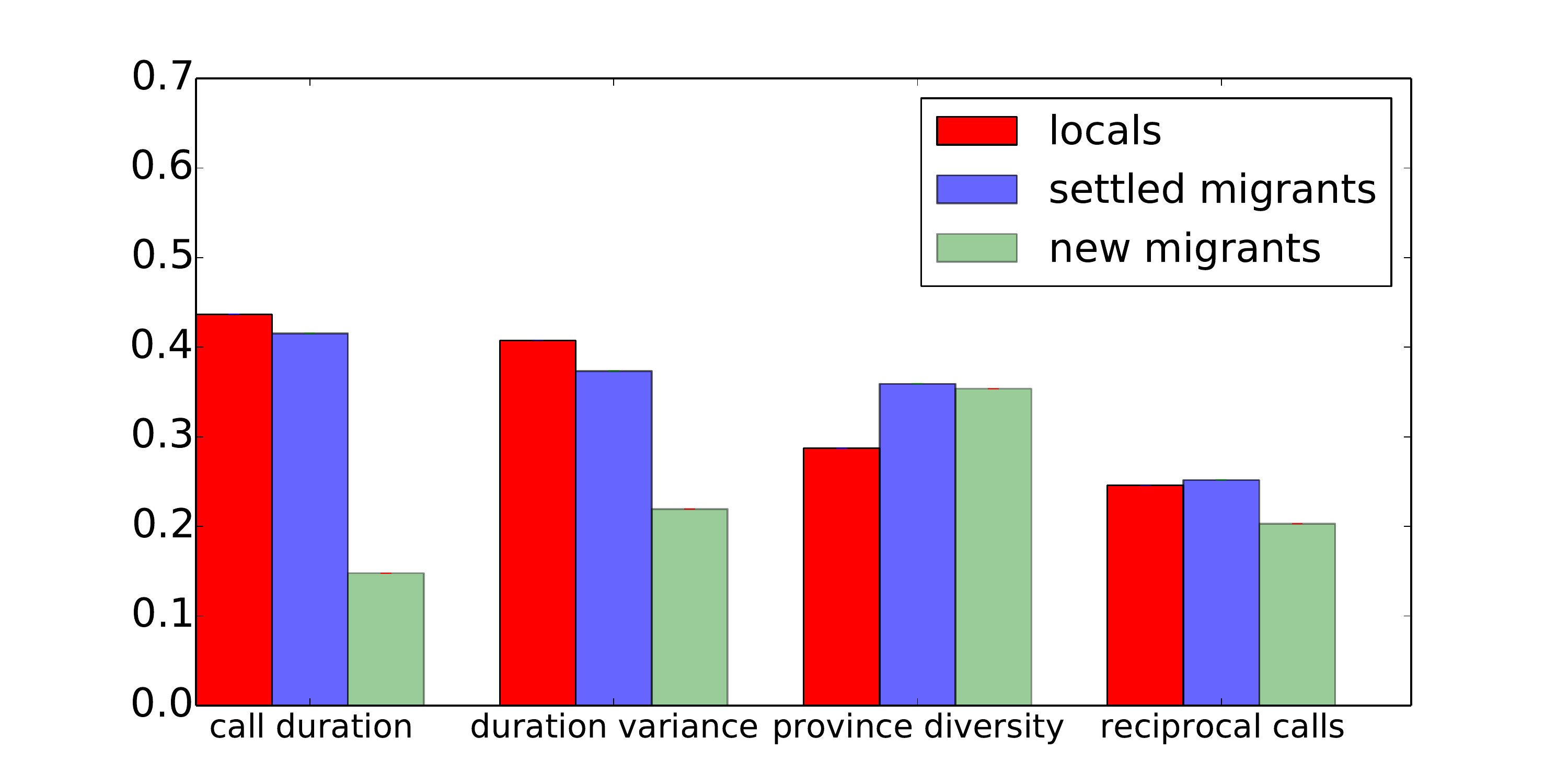, width=0.4\textwidth}
}
%
\subfigure[Geographical features.\label{fig:feature_geo}] {
\epsfig{file=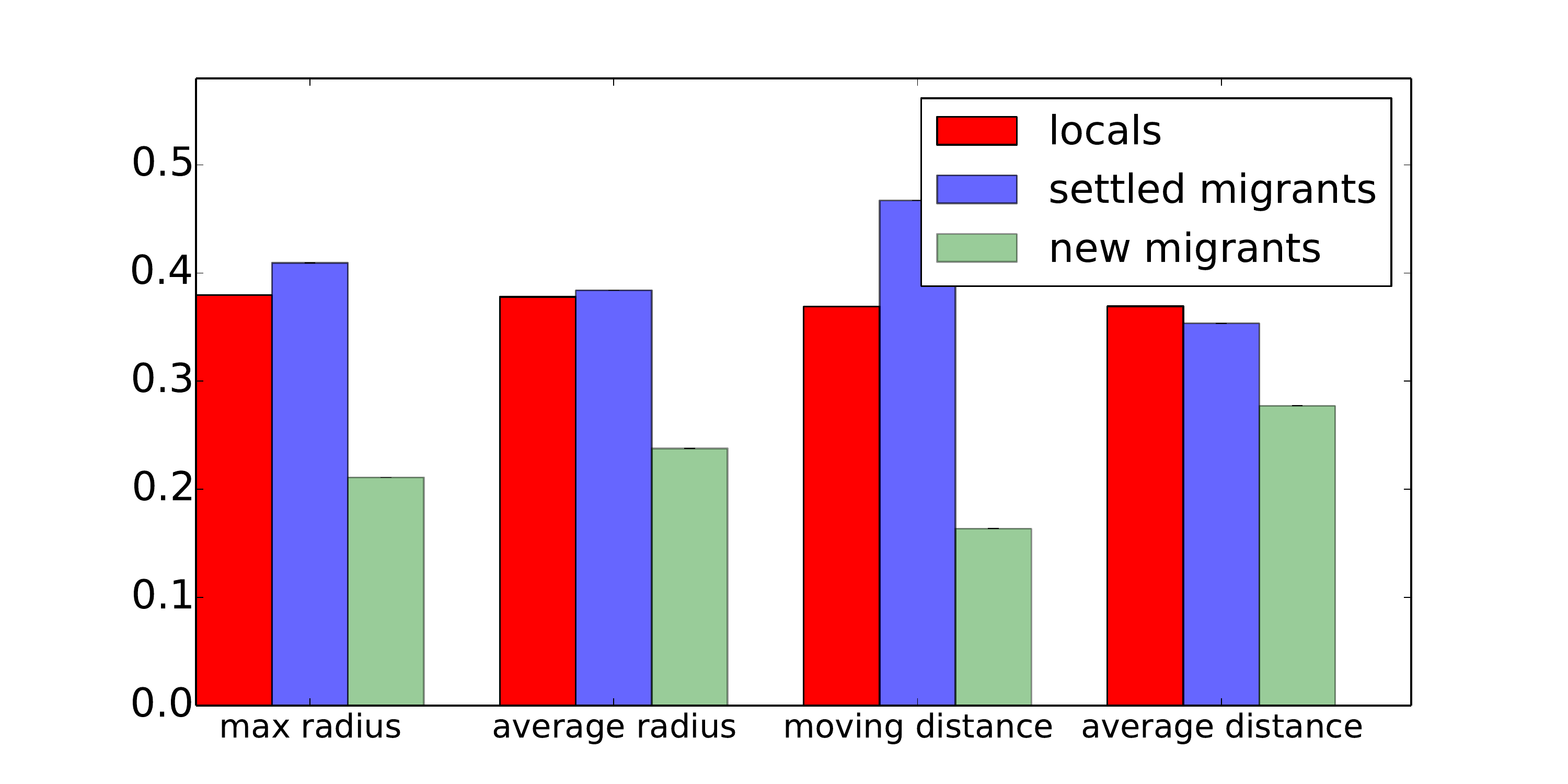, width=0.4\textwidth}
}
\caption{Feature comparison between \locals, \migrants and \newcomers.
Different colors represent different groups of people.
Since different features may end up in very different scales, we normalize each feature group in this figure so that their means sum up to 1, except for demographics of contacts and fraction of reciprocal calls (they all naturally fall between 0 and 1).
Error bars represent standard errors, and they are tiny.
} \label{fig:feature_comparison}
\end{figure*}

\para{Mobile communication networks.} One core component of our study is a weekly mobile communication network based on the call logs.
Grouping by weeks allows us to account for variations between weekdays and weekends.
Formally, we build a directed graph $G_t=(V_t, E_t)$ for each week $t$ ($t \in \{1, 2, 3, 4\}$), where $V_t$ is the set of users, and each directed edge $e_{ij} \in E$ indicates that $v_i$ calls $v_j$ ($v_i, v_j \in V_t$). 
Note that only a subset of users in $V_t$ are labeled as \locals, \migrants or \newcomers (in total around 3 million users).
This subset is the focus of our study.

\para{Geographical locations.} Another component is the geographical locations that a person is active at.
Specifically, for each call a person makes, we have access to the GPS location from the corresponding telecommunications tower.
We use each week as a window and collect all the locations that a person makes calls in that week, and refer to this ordered list of locations for user $v$ at week $t$ as $L_v^t=[l_1, \ldots, l_n]$, where $l_i$ contains the latitude and the longitude.
We have geographical locations for the subset of users with labels since they are all users of China Telecom by definition.

We will computationally characterize these two components using features in \secref{sec:analysis}.

\section{\Locals, Settled Migrants, New Migrants}
\label{sec:analysis}

To understand how \locals, \migrants and \newcomers differ from each other,
we examine a wide range of features from people's mobile communication networks and geographical locations.
To observe the initial state of urban migrants without much integration, 
we use the data from the first week after \newcomers joined China Telecom (week 2) in this section.
We will focus on the integration process in \secref{sec:integration},
in which we also show that most features do not change for \locals and \migrants in future weeks.
\tableref{tb:feature} provides an overview of the features that we consider.
Demographics of users' friends,\footnote{Note that although we do not have demographics information for users using other service providers, we have demographics information for a much larger set than the ones labeled \locals, \migrants or \newcomers because we require these three groups to be active in each week.}
ego-networks features, and call behavior derive from the mobile communication networks, while geographical features come from location information.
In the following, we will explain the motivation and related theories of each feature.

\para{Demographics of contacts (\figref{fig:feature_attribute}).}
A person's mobile communication network can reasonably approximate her social network.
\Locals likely maintain very different social networks from migrants since they have grown up in this city.
Also, as a person settle down in a new city, her social network may change dramatically.
Existing studies suggest that kin relationships play an important role in determining the destination of migration \cite{gurak1992migration} and relationship with \locals are crucial for migrant integration \cite{Liu:HabitatInternational:2012}.
We look at the demographics of contacts in age, sex, and birthplaces.

{\noindent \bf \em Homophily in sex and age.}
It is well recognized that people tend to make friends with those who are similar to themselves, also known as homophily \cite{mcpherson2001birds}.
We observe interesting contrasts regarding homophily of age and sex.
\Locals show the strongest homophily in sex, i.e., \locals have the largest fraction of contacts with the same sex.
Surprisingly, in terms of the absolute fraction, \newcomers have 
more contacts with a different sex than with the same sex (only around 40\%).
In contrast, \locals are less likely to have contacts at similar ages than \migrants, but more than \newcomers.

{\noindent \bf \em Birthplaces.}
The most striking difference lies in that 70\% of a local's contacts are also \locals.
This number is much lower for \migrants, and the lowest for \newcomers.

Townsmen, people who share the same hometown (exclude Shanghai), are an important component of a \newcomer's initial network (30\% of \newcomers' contacts are townsmen).
This observation echoes existing findings regarding kin relationships. 
In comparison, \migrants have an even higher fraction of townsmen in their contacts, which suggests that \newcomers get to know more people from the same hometown as they integrate into a city.
These observations are consistent with homophily, but they also indicate that urban migrants maintain a relatively separate personal network from \locals.

\para{Ego-network characteristics (\figref{fig:feature_network}).}
As expected, \newcomers have the smallest degree and weighted degree.
However, \migrants tend to have the largest degree, larger than \locals.
This indicates an interesting transition that migrants may undergo.
Maybe because of homophily, neighbors of \migrants also have the largest average degree, and neighbors of \newcomers have the smallest degree.

Clustering coefficient measures the fraction of triangles in the ego-networks.
It roughly reflects how connected a person's contacts are to each other.
Interestingly, \newcomers present the largest clustering coefficient, while \migrants have the lowest.
It may suggest that \newcomers start with a close-knit group when they move to a big city like Shanghai.
Connecting with our previous observations, this close-knit group tend to come from the same province as the \newcomers.
It is worth noting that this could also relate to that \newcomers have the smallest ego-networks.

\begin{figure*}[t]
\centering
\begin{tabular}{cccc}
\multicolumn{4}{c}{\large \newcomers move towards \migrants and both move towards \locals} \\[-0.08in]
\subfigure[Same-sex.\label{fig:weekly_sexnumhis}] {
\epsfig{file=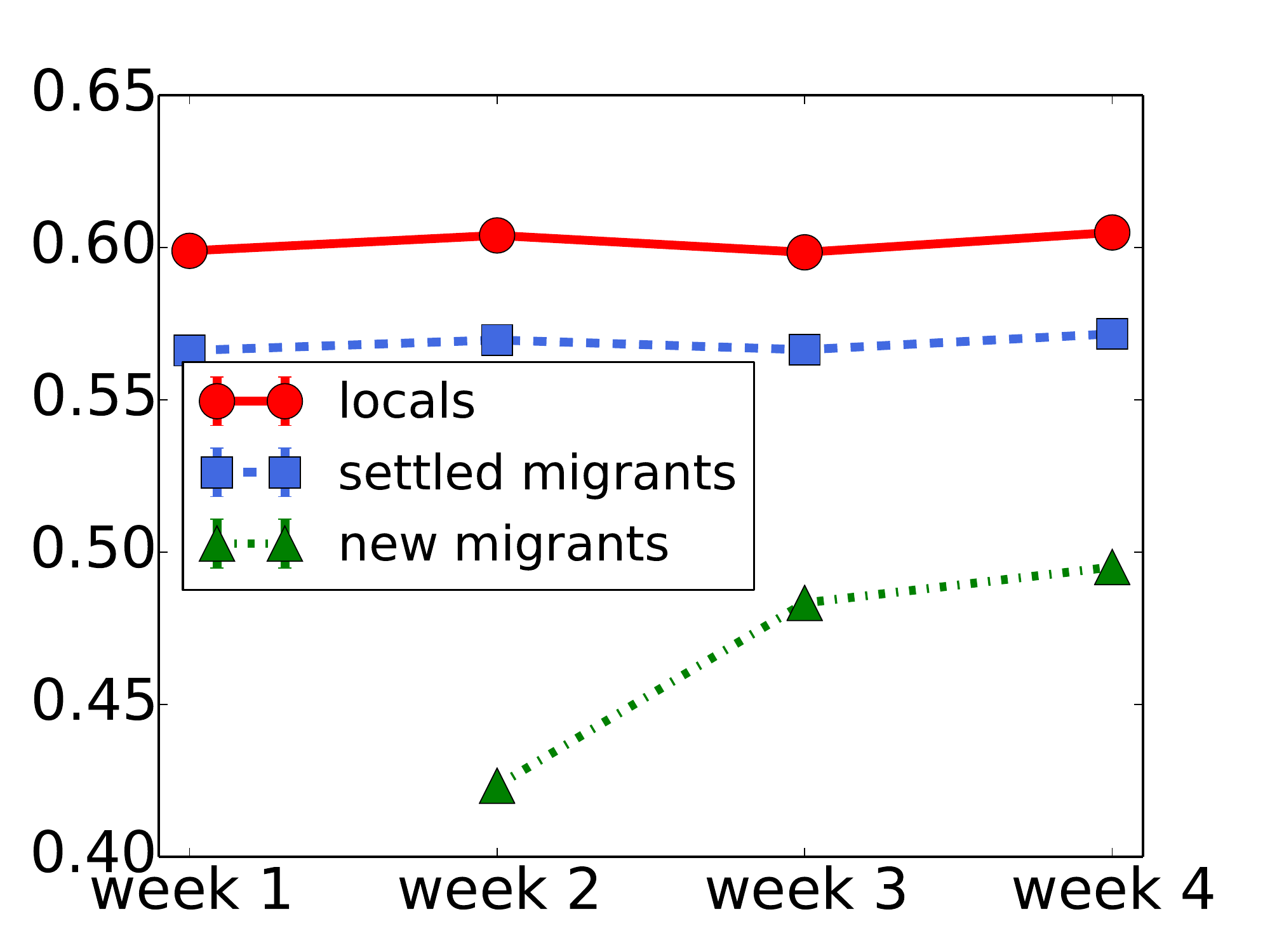, width=0.23\textwidth}}
&
\subfigure[Local.\label{fig:weekly_callsh}] {
\epsfig{file=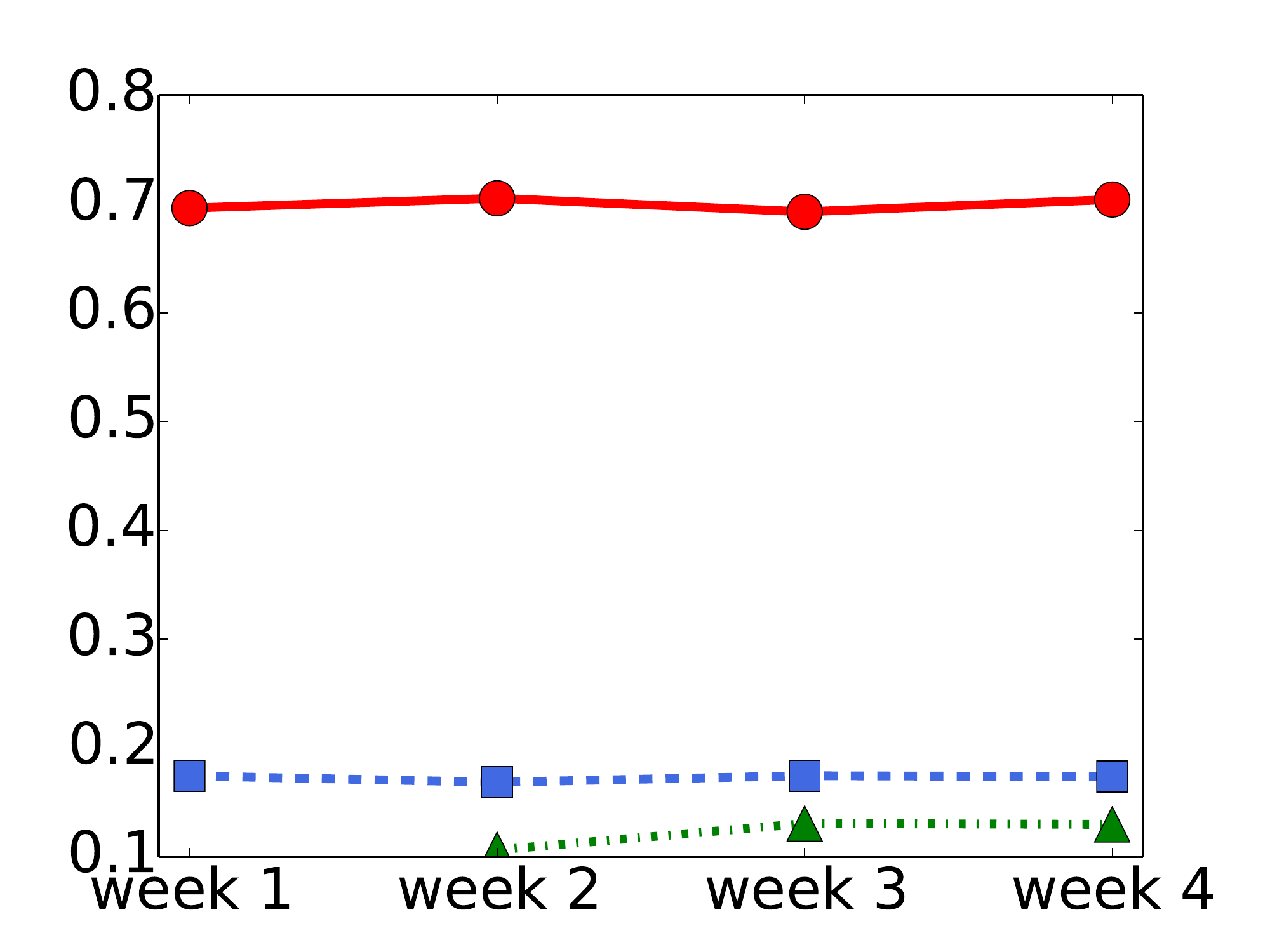, width=0.23\textwidth}}
&
\subfigure[Call duration (in seconds).\label{fig:weekly_callduration}] {
\epsfig{file=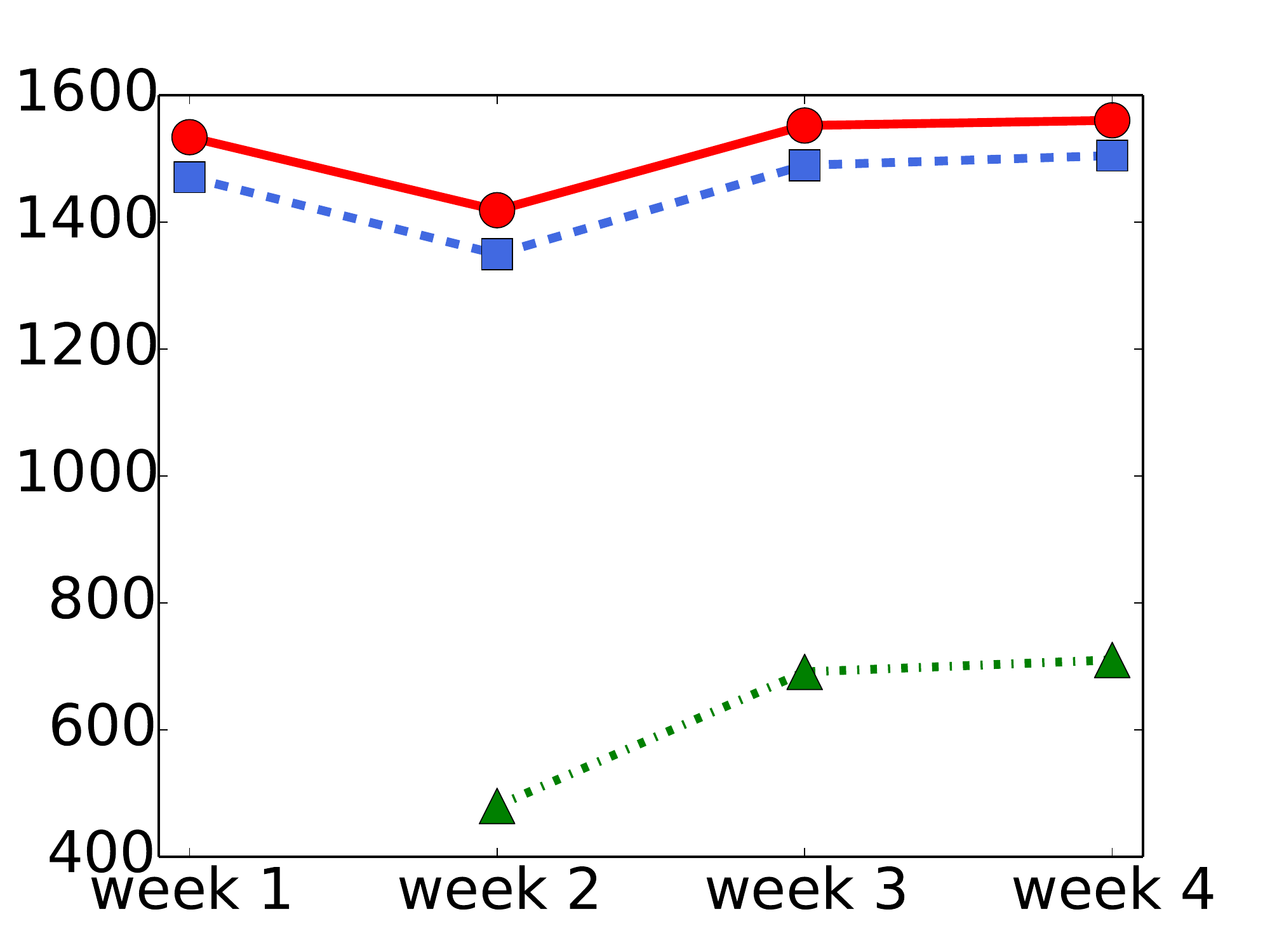, width=0.23\textwidth}}
&
\subfigure[Average moving distance.\label{fig:weekly_range_ave}] {
\epsfig{file=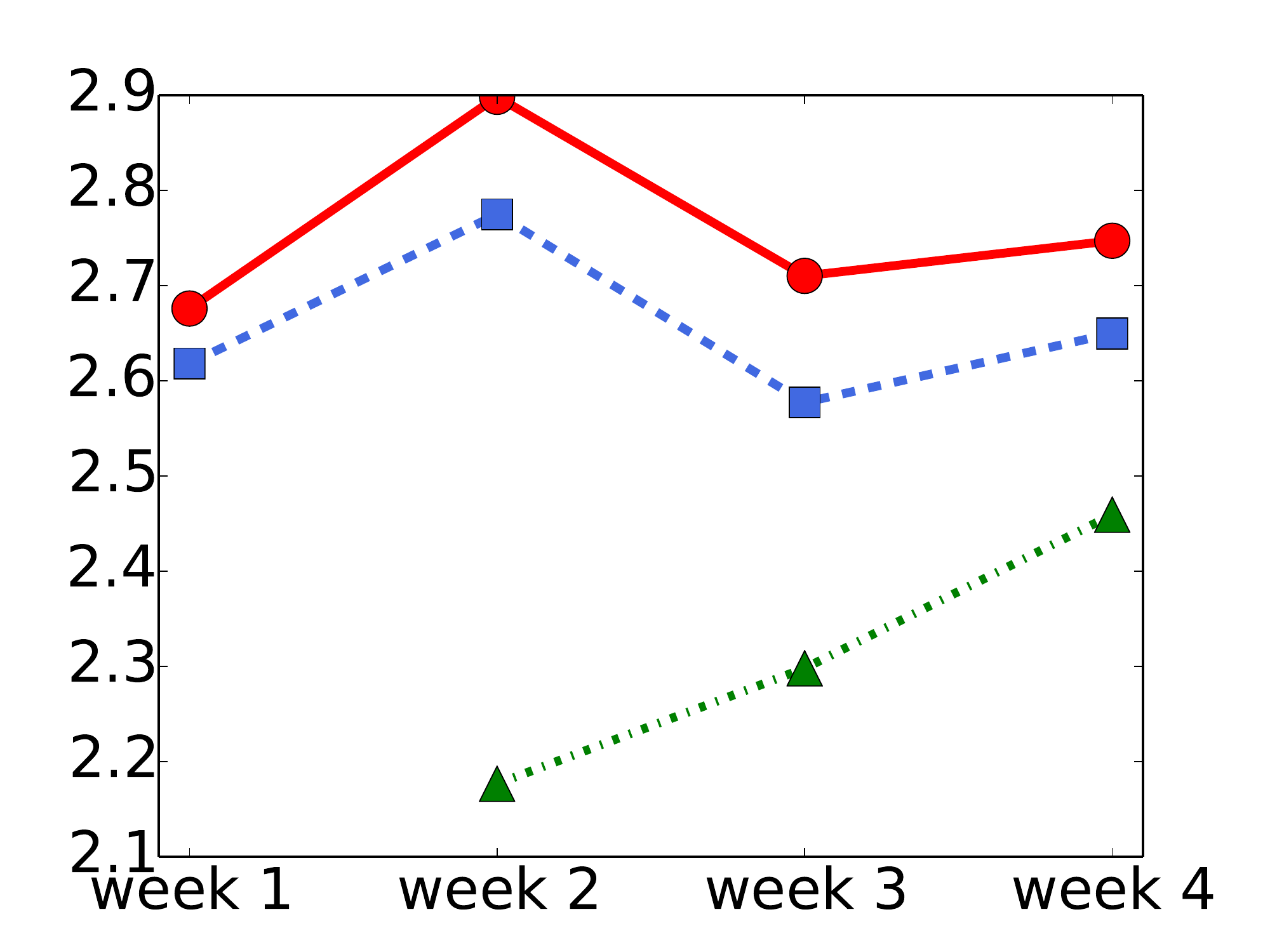, width=0.23\textwidth}}
\\
\midrule
\multicolumn{4}{c}{\large \newcomers move towards \locals initially, but will likely eventually move away from \locals} \\[-0.08in]
\subfigure[Degree.\label{fig:weekly_degree}] {
\epsfig{file=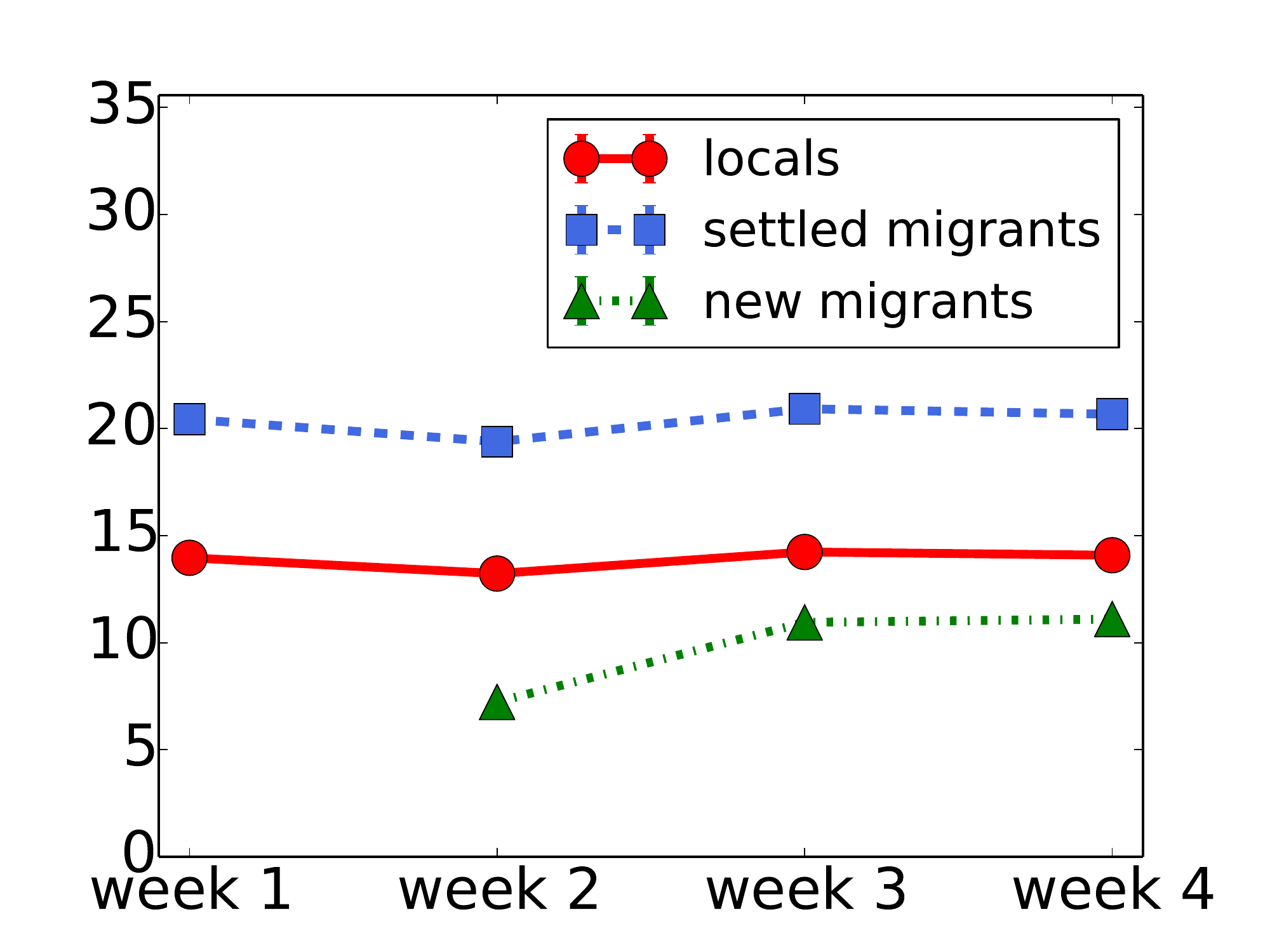, width=0.23\textwidth}}
&
\subfigure[Clustering coefficient.\label{fig:weekly_cc}] {
\epsfig{file=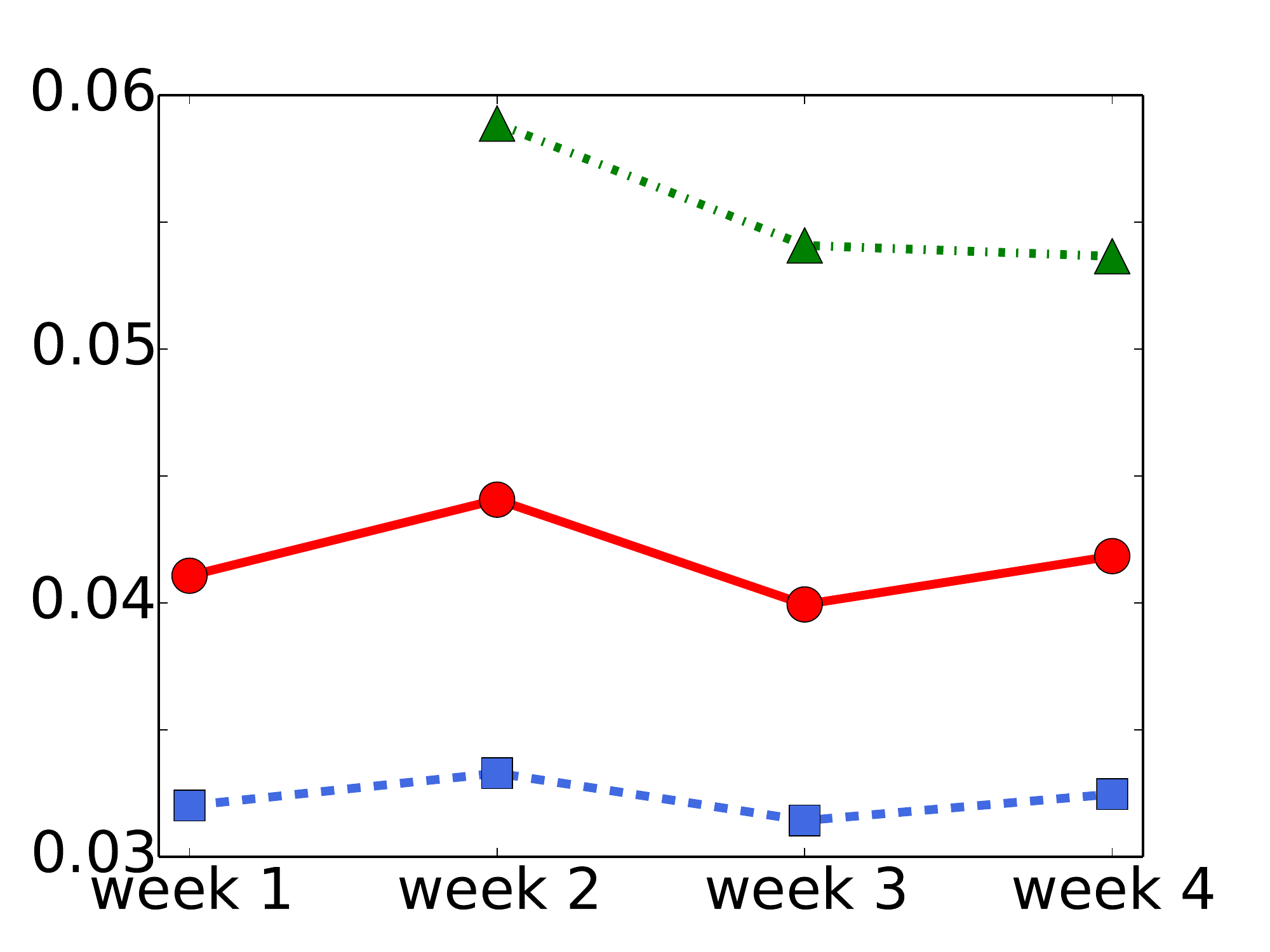, width=0.23\textwidth}}
&
\subfigure[Reciprocal calls.\label{fig:weekly_bd}] {
\epsfig{file=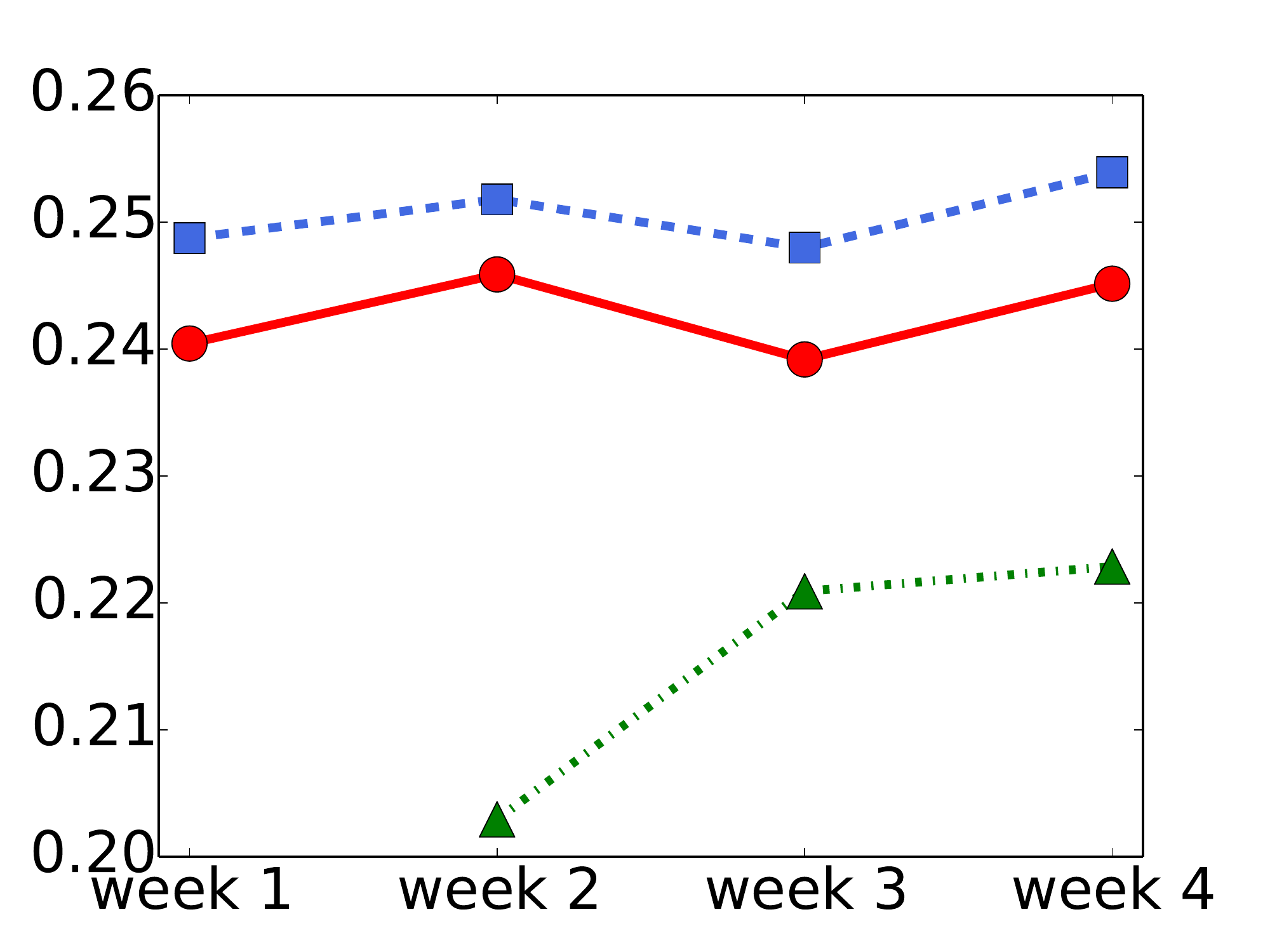, width=0.23\textwidth}}
&
\subfigure[Max radius.\label{fig:weekly_gyrationradiusmax}] {
\epsfig{file=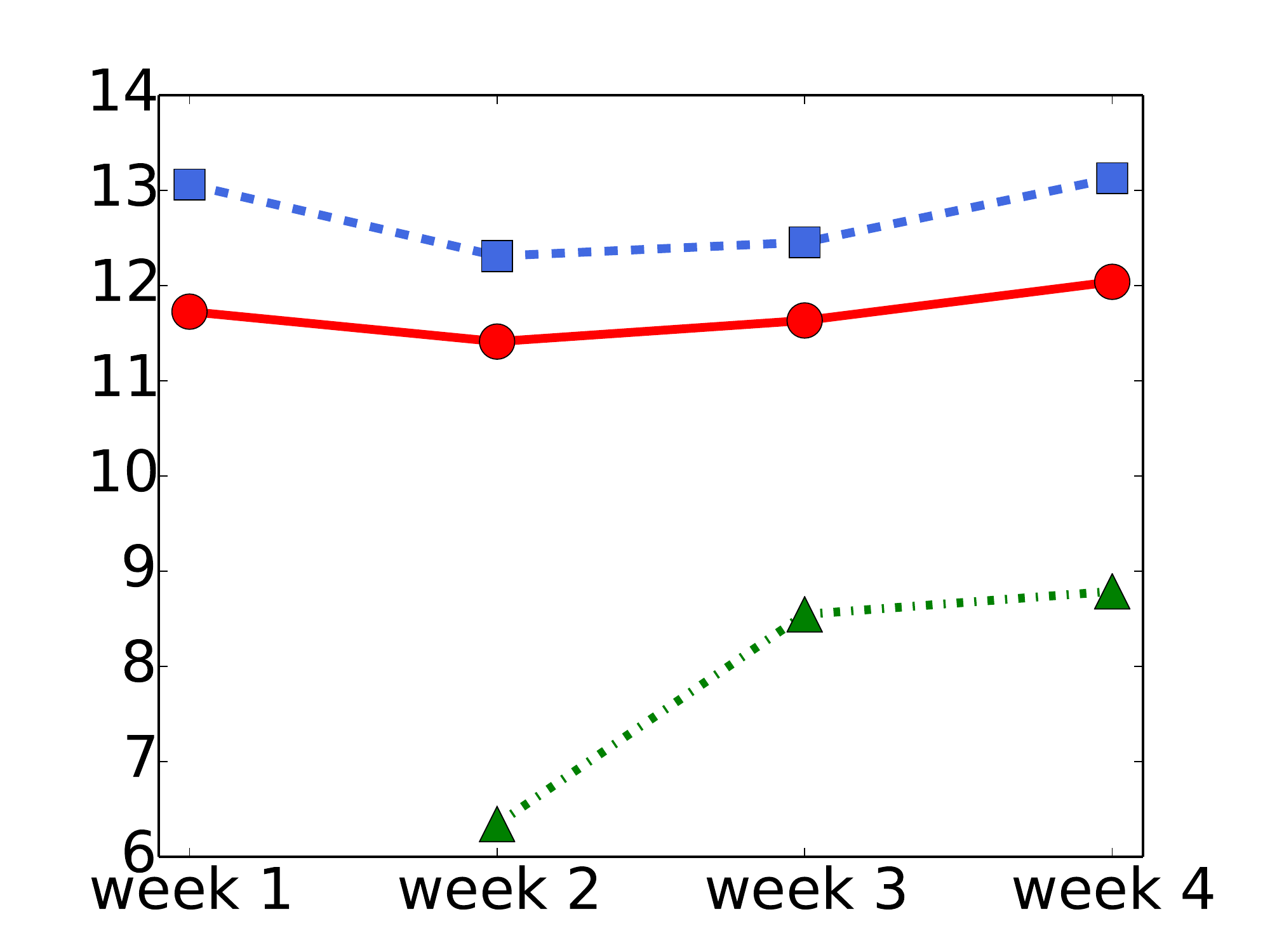, width=0.23\textwidth}}
\\
\midrule
\multicolumn{4}{c}{\large \newcomers move towards \migrants but away from \locals} \\[-0.08in]
&
\subfigure[Province diversity.
\label{fig:weekly_provdiv}]{
\epsfig{file=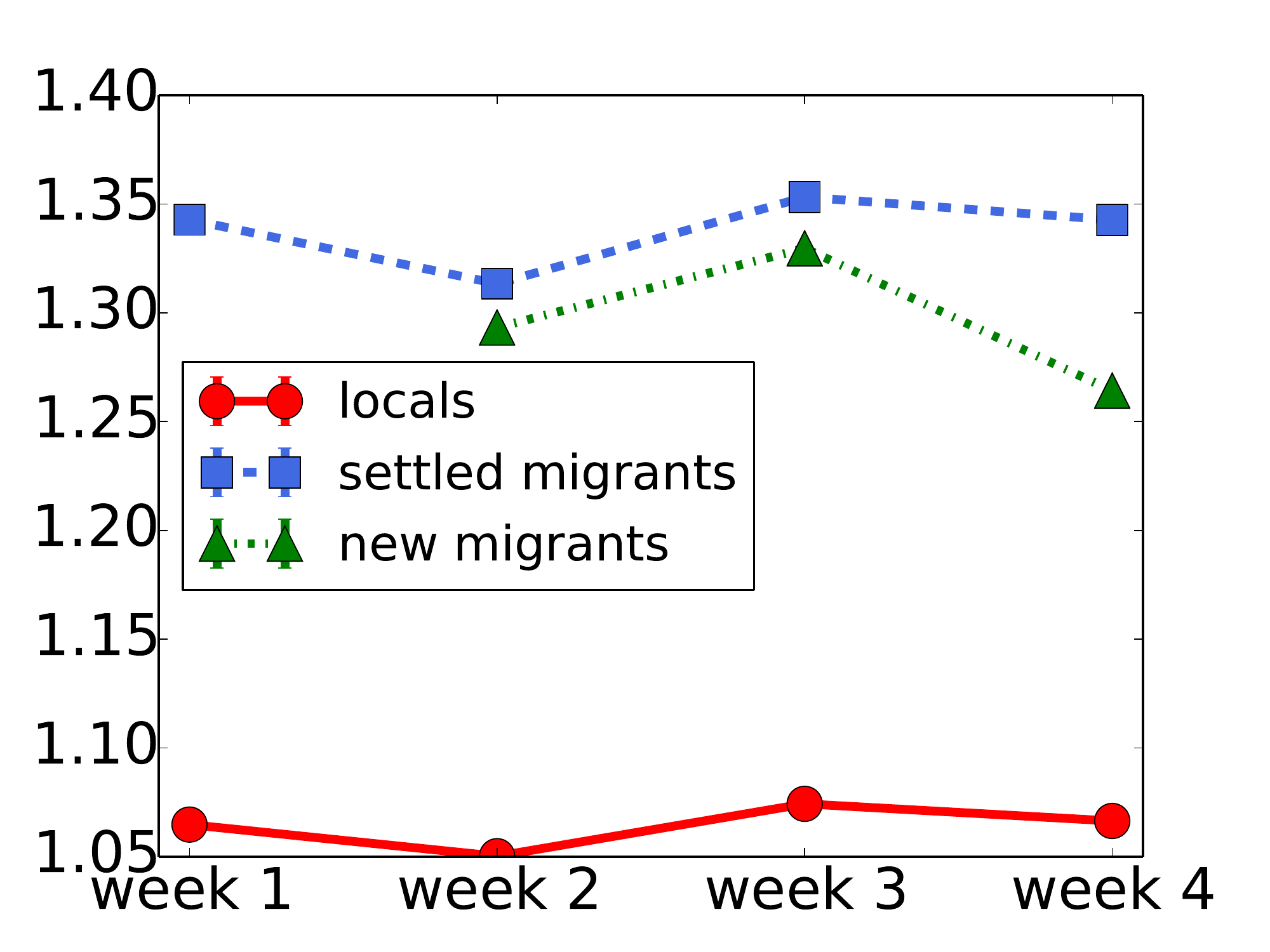, width=0.23\textwidth}}
&
\subfigure[Fraction of townsmen.\label{fig:weekly_townsman}]{
\epsfig{file=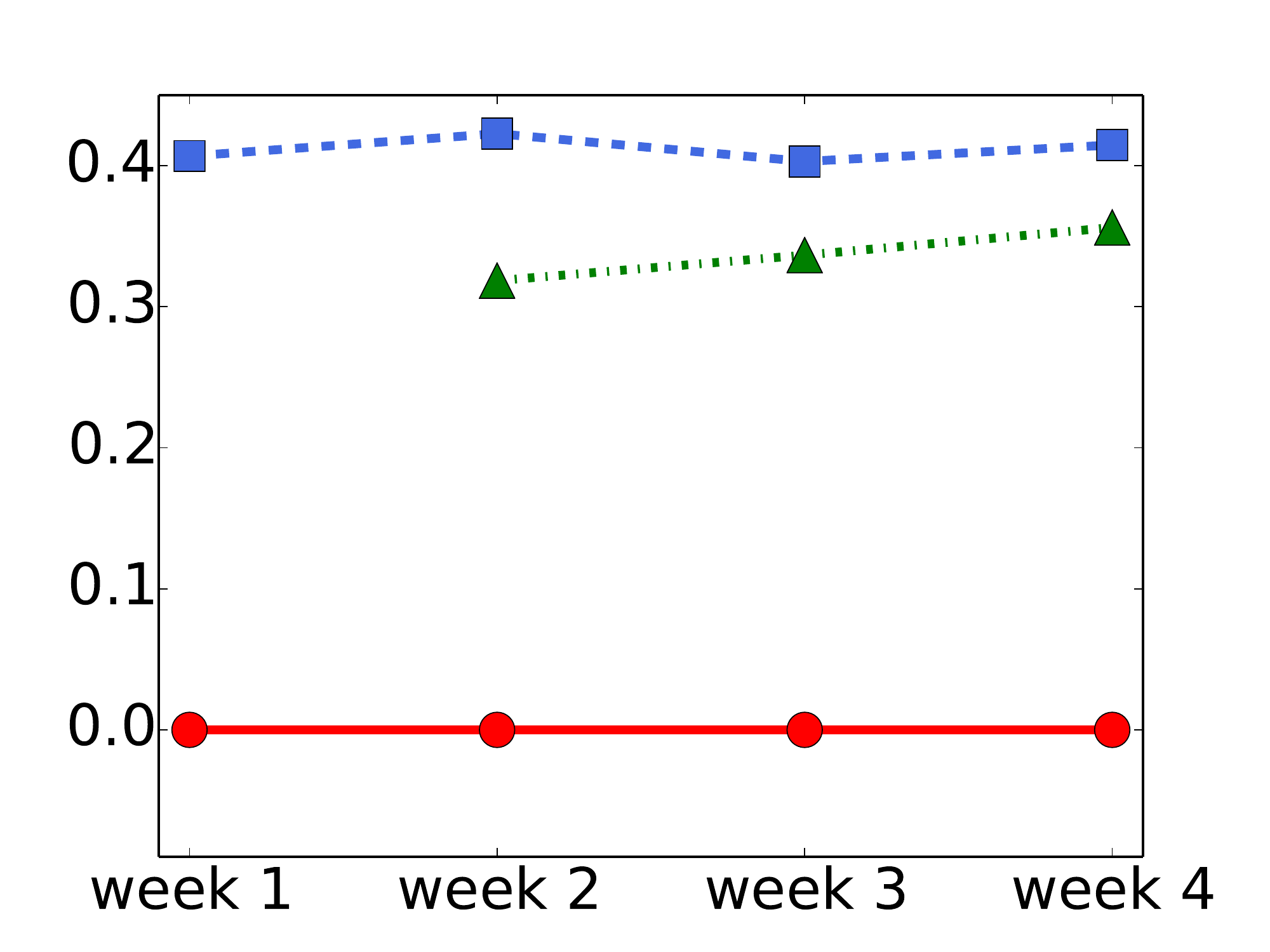, width=0.23\textwidth}}
&
\end{tabular}
\caption{Integration process of \newcomers. 
Each figure presents how the values of a feature evolves over the four weeks for \locals, \migrants, and \newcomers. 
We choose four samples in the first two rows because there are more than four features that belong to those categories.
Error bars represent standard errors.
}\label{fig:integration}
\end{figure*}

\para{Calling behavior (\figref{fig:feature_call}).} The duration of calls reflects the nature of relations between a person and her contacts.
Calls of long duration likely involve intimate relations or are driven by substantial businesses, while calls of short duration tend to be quick check-ins or relate to small incidences.
We find that \locals and \migrants have similar levels of average call duration, much larger than \newcomers.
Similar trends show up in the variance of call duration.

Regarding the diversity of provinces in a person's contacts, \migrants have the most diverse group of contacts, while \locals have the lowest.
This pattern resonates with previous observations that \locals have about 70\% of contacts that are also \locals.

Finally, we find that \locals and \migrants are more likely to %
have reciprocal relationships with their contacts, while the fraction of reciprocal calls is the lowest for \newcomers.
This again shows that the personal networks of \newcomers are still nascent.
Note that the difference is much less dramatic than that in call duration.

\para{Geographical patterns (\figref{fig:feature_geo}).}
The mobility of people in different groups can be reflected by their locations over time.
As we have discussed in the introduction, both \migrants and \newcomers tend to move around the central part of Shanghai, while \locals are more disproportionally frequent in the periphery.
Regarding the radius of a person's movement around her own center, we observe that \migrants have the largest radius both in terms of max radius and average radius.
This suggests that although \newcomers start with a smaller active area than \locals, \migrants move in an even larger area than \locals.

Total moving distance is correlated with the total number of calls that a person makes.
We thus discover the same ordering as in weighted degree.
However, \locals tend to move the most distance between calls on average, while \newcomers move the shortest distance.
This further suggests that \newcomers have a smaller active area than \locals.

\para{Summary.}
Comparing \migrants to \locals, we observe that \migrants have more active and diverse behavior patterns both in mobile communication networks and in geographical movements.
Meanwhile, \newcomers present different characteristics from both \migrants and \locals.
This suggests that \newcomers go through significant changes in their communication networks and geographical locations as they slowly 
settle down.

\section{Integration of \NEWCOMERS}
\label{sec:integration}

Given the differences between \locals, \migrants and \newcomers that we have observed, we now investigate the integration process of \newcomers.
Since a subset of \newcomers eventually become \migrants,
we hypothesize that the features of \newcomers will grow more similar to those of \migrants in week 3 and week 4.
Indeed, we find that \newcomers are slowly ``becoming'' \migrants in most features.
\figref{fig:integration} presents how some features of \locals, \migrants and \newcomers change over the four weeks (\newcomers only moved to Shanghai in week 2).
Although existing studies have argued that different generations of migrants can exhibit different characteristics \cite{portes2002price,chiswick2004educational}, our observation shows that the features that we propose are robust to generation differences, or Shanghai is too young a city to observe generation gaps from telecommunication records.

The more interesting comparison is with \locals.
One possible way to evaluate migrant integration is whether they become more similar to \locals over time.
Depending on how the features of \locals compare to \newcomers and \migrants, we can observe several possible trajectories as shown in
\tableref{tab:ordering}.
An ideal integration process suggests that \newcomers become more similar to \locals, and \migrants represent a middle state in this process, i.e.,
the orderings should follow $\textit{locals} > \textit{settled} > \textit{new}$ or $\textit{locals} < \textit{settled} < \textit{new}$,
and we should observe that the features of \newcomers move towards \migrants in week 3 and week 4.
Some features indeed show consistent trajectories with this ideal integration process, including
fraction of same-sex contacts, fraction of local contacts, call duration, duration variance, and average moving distance.
It makes sense that migrants are probably never going to match \locals in the fraction of local contacts, but such matching may happen in average moving distance and call duration.

However, for the majority of features,
we observe that although \newcomers initially move towards \locals, they may eventually become further away from \locals after settling down.
These features include degree, weighted degree, average degree of neighbors, clustering coefficient, fraction of friends with similar age, fraction of reciprocal calls, max radius, average radius, total moving distance.
In particular, all features in ego-network characteristics follow this trajectory, suggesting that \newcomers eventually build quite different communication networks from \locals.

It is rare that \newcomers do not move towards \locals at all but become more different from \locals in the integration process.
This only happens in province diversity and townsmen.\footnote{It is tricky for townsmen, since \locals do not have townsmen that are not from Shanghai and always have 0 in this feature.}
Both features point to the fact that \newcomers start with a more diverse communication network in terms of birthplaces than \locals, and their networks get even more diverse over their stay in Shanghai.
Note that there is a decline in province diversity for \newcomers in week 4 but they are still closer to \migrants than to \locals.

\begin{table}[!t]
\small 
\begin{tabular}{p{0.17\textwidth}|p{0.27\textwidth}}
\toprule
$\text{locals} > \text{settled} > \text{new}$, $\text{locals} < \text{settled} < \text{new}$ & New migrants are moving towards \locals and \migrants are in the middle of this process.\\
\midrule
$\text{new} > \text{locals} > \text{settled}$, $\text{new} < \text{locals} < \text{settled}$ & New migrants move towards \locals initially, but eventually move away from them and remain different from \locals after they settle down.\\
\midrule
$\text{locals} > \text{new} > \text{settled}$, $\text{locals} < \text{new} < \text{settled}$ & Settled migrants and \locals are different, and \newcomers never move towards \locals.\\
\bottomrule
\end{tabular}
\caption{All possible orderings of feature values between \locals, \migrants, and \newcomers.
}
\label{tab:ordering}
\normalsize
\end{table}

Interestingly, in some features, we observe that the integration slows down or converges in week 4 for \newcomers. 
This is likely due to the fact that not all \newcomers are going to become \migrants.
As a result, we can already see that the integration process stops or slows down in week 4 for a subset of these people.

\para{Discussion.}
Overall, we find that \newcomers are settling down and gradually becoming \migrants, and this observation is robust with potential generation gaps.
However, 
in a substantial fraction of the features, 
although \newcomers are temporarily moving towards \locals, they are probably going to become different from \locals as \migrants do.
In other words, 
despite settling down, \migrants remain fairly different from \locals.

\section{Distinguishing Migrants from Locals}
\label{sec:prediction}

\begin{figure*}[t]
\vspace{-0.2in}
\centering
\subfigure[Prediction performance in distinguishing \migrants from \locals. \label{fig:res_taskone}] {
\epsfig{file=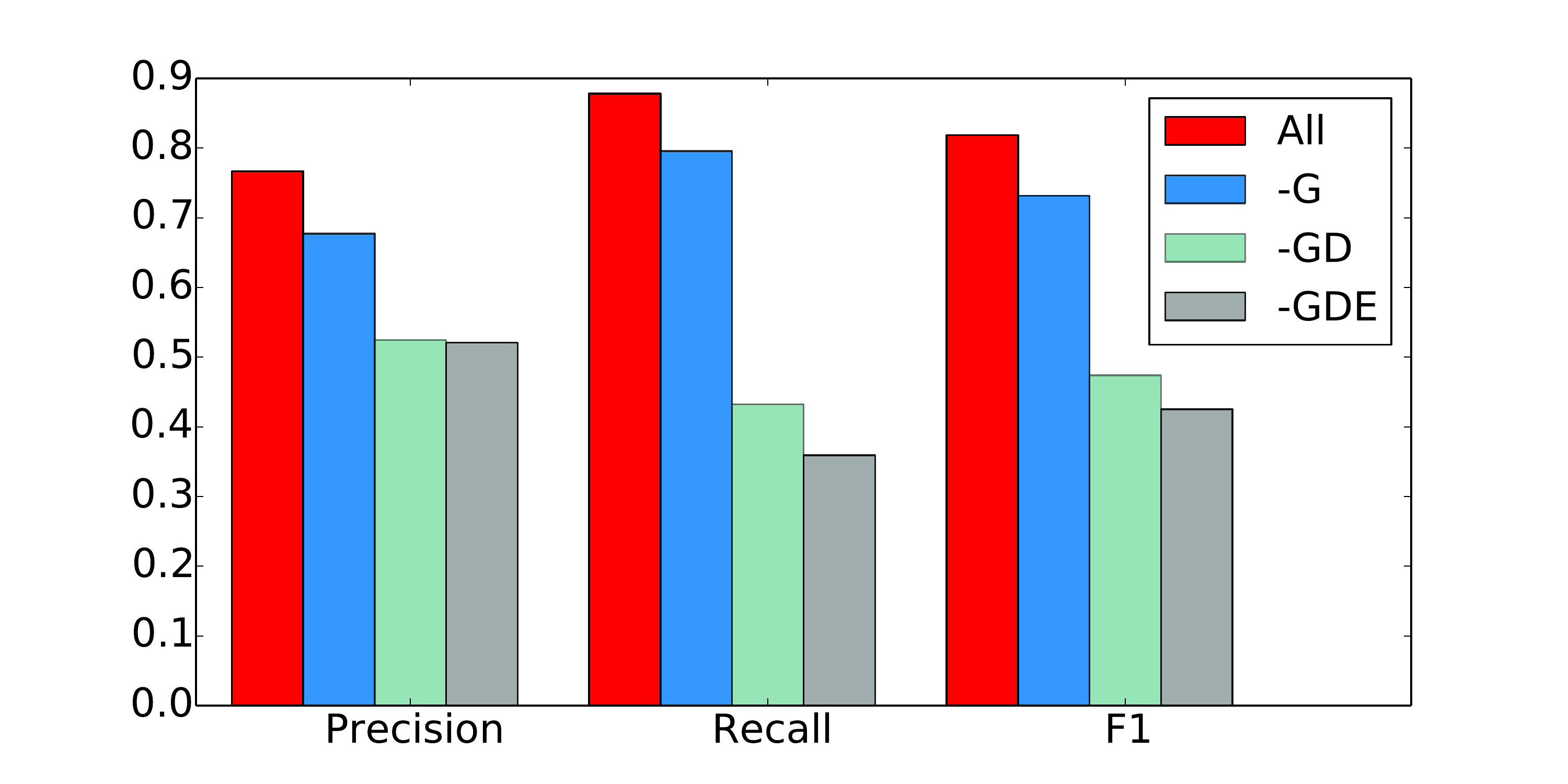, width=0.45\textwidth}
} 
\hspace{-0.15in}
\hfill
\subfigure[Fraction of migrants classified as locals. \label{fig:local_rate}] {
\epsfig{file=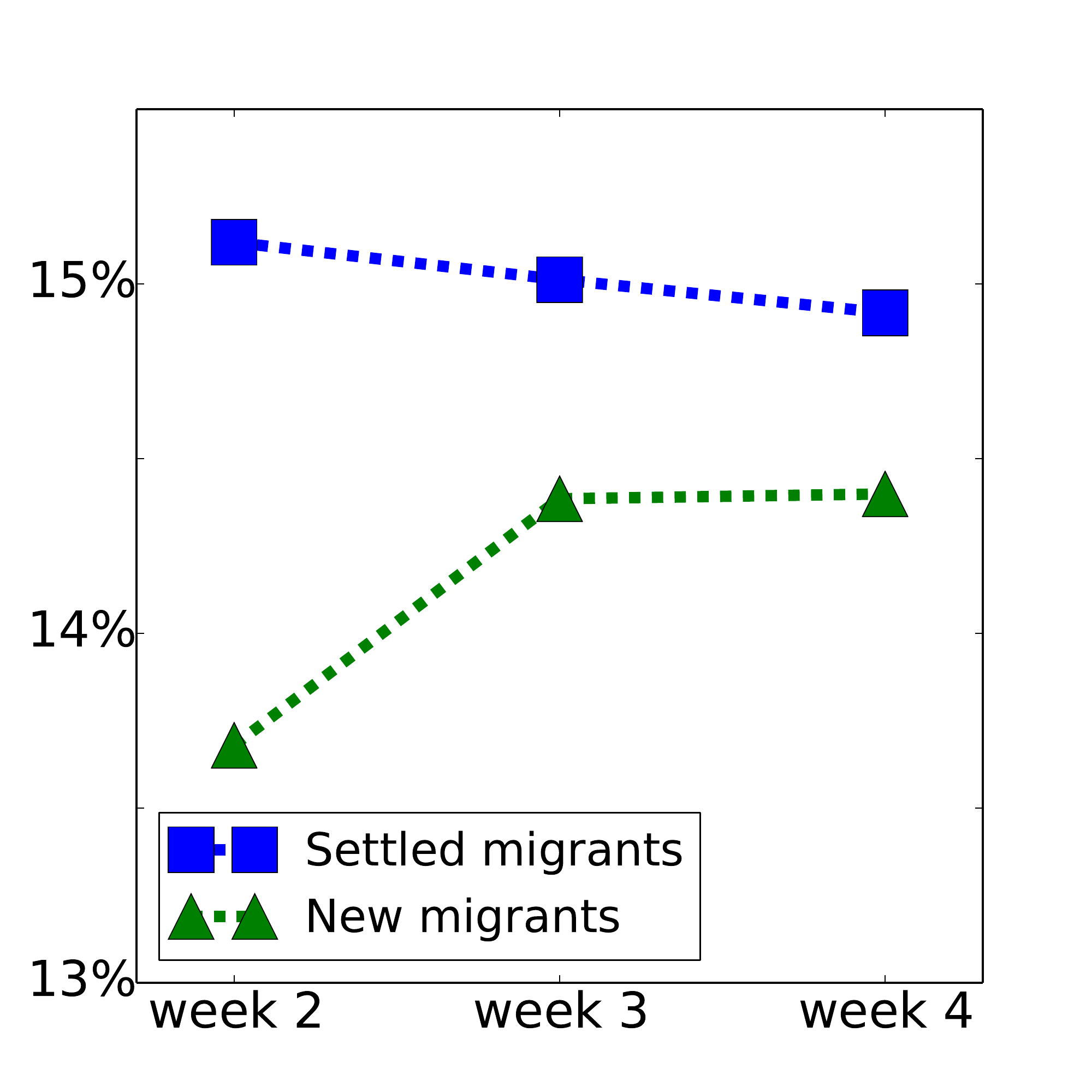, width=0.23\textwidth}
}
\hspace{-0.15in}
\hfill
\subfigure[Performance of three-way classification. \label{fig:recall_weekly}] {
\epsfig{file=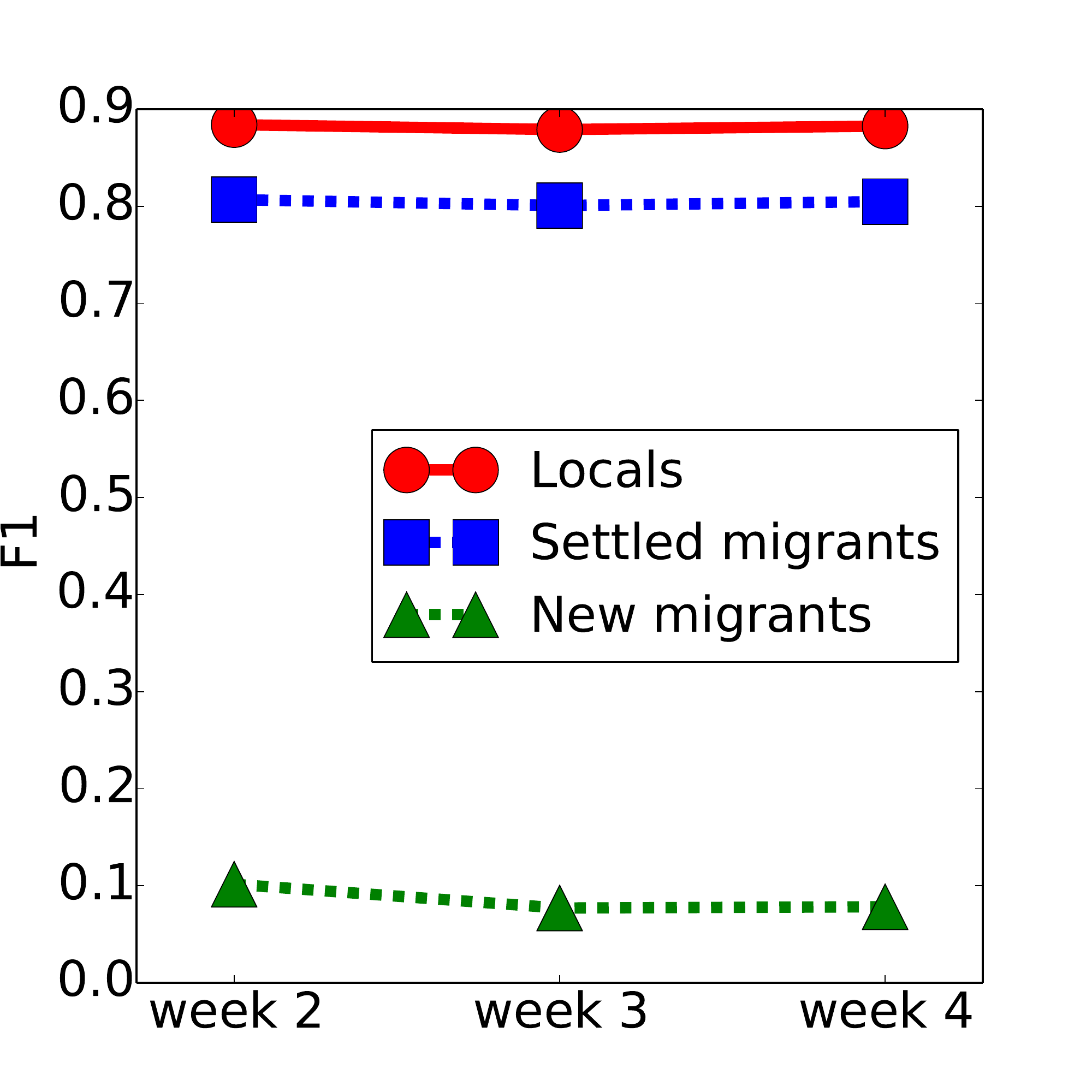, width=0.23\textwidth}
}
\caption{Performance of distinguishing migrants.
\small
\figref{fig:res_taskone} and \figref{fig:local_rate} are results from the binary classification between \migrants and \locals.
\figref{fig:res_taskone} shows the performance of feature ablation (we show the feature class that is the most influential in each ablation step),
while \figref{fig:local_rate} shows the fraction of \migrants and \newcomers that is mistakenly classified as \locals over time.
\figref{fig:recall_weekly} shows the F1 scores in the three-way classification problem over time.
\normalsize
\label{fig:performance_weekly}} 
\end{figure*}

We set up two prediction tasks to assess the difficulty of distinguishing migrants from \locals with the features that we propose.
Since the number of \newcomers is much smaller compared to \migrants and \locals (22K vs. 1.0M and 1.7M), we employ two formulations in this section.
First, we propose a binary classification task to distinguish \migrants from \locals.
We then apply this binary classifier to \newcomers to evaluate how often a \newcomer would be mistakenly considered as a \local. %
This misclassification rate can reflect how well \newcomers have integrated, at least in terms of fooling our classifier.
Second, we work on the more challenging three-way classification problem to identify \newcomers, \migrants, and \locals.

\vpara{Experiment setup.}
In both prediction tasks,
each instance consists of features based on a user's calling logs within one week. 
We randomly draw $50\%$ of users and use their calling logs in week 2 to train the classifier.
The remaining data is used to test the classifier ($50\%$ of data in week 2, and $100\%$ of data in week 3 and week 4). 
In particular, we use all features listed in Table~\ref{tb:feature} except ``townsmen'', as measuring the fraction of townsmen relies on the user's label (the user's birthplace).
We use precision, recall, and F1-score for evaluation, with the minority class (i.e., migrants) as the target class. 
For the classifier, we use $\ell_2$-regularized logistic regression.
We choose the best $\ell_2$ penalty coefficient using 5-fold cross-validation in training data.

\subsection{Settled Migrants vs. Locals %
}

\vpara{Prediction performance (\figref{fig:res_taskone}).}
It turns out to be relatively easy to distinguish \migrants from \locals.
We can achieve an F1-score of 0.82 with all the features that we propose.
We further analyze the contribution of each type of features by removing them one by one.
In the first removal step, we find that
geographical features were the most influential feature set,
i.e., F1 drops the most (0.11) if we remove geographical features ({\em -G}). 
Demographics is the most important in the second step ({\em -GD}).
In the third step, removing ego-network features is the choice,
leaving us with a classifier that only uses call behavior ({\em -GDE}).
F1 drops 
almost 50\% after removing these three types of features. 
In addition, the prediction performance of the classifier is robust over time: F1-scores on each week vary little ($<$0.0007). %

\vpara{Integration of \newcomers (\figref{fig:local_rate}).} 
One way to evaluate the integration of \newcomers is to measure how often this binary classifier would mistakenly classify a \newcomer as a \local.
We present the fraction of  misclassified \locals among \migrants as a comparison point.
Overall, \migrants are more likely to be misclassified as \locals than \newcomers (e.g., 15.2\% vs 13.6\% in week 2). 
However, we observe an increasing trend for \newcomers over the three weeks.
In week 3, the misclassified fraction of \newcomers increases to 14.4\%, suggesting that they become more similar to \locals over time. 
The growth slows down in week 4, which is consistent with our findings in \secref{sec:integration}.
To our surprise, the faction of \migrants misclassified as \locals slightly decreases over time. 
This suggests that some \migrants could have stopped integrating 
with \locals after settling down, but build their own communities and keep their own lifestyles instead.
\subsection{Identifying New Migrants %
}

The three-way classification problem is challenging
due to the relatively small number of \newcomers (about $0.8\%$ of all instances).
The classifier only achieves an F1-score of $0.1$ on identifying \newcomers and this performance drops over time, while the performance on \migrants and \locals remains similar to the binary classification task (\figref{fig:recall_weekly}).
We find that more \newcomers are classified as \migrants or \locals incorrectly by the classifier over time.
This is consistent with the observation that \newcomers are becoming similar to \migrants or \locals in most characteristics despite the short time span, while \migrants and \locals tend to stay constant.

\section{Related Work}
\label{sec:related}

Migrant integration is a well-recognized research question in many disciplines.
Most relevant to our work is the study of urban migration \cite{brown1970intra,schiller2009towards,Fischer:ToDwellAmongFriendsPersonalNetworks:1982,schiller2011locating,scholten2013agenda,brockerhoff1995child,whitzman2006intersection,glaeser2001cities,Goodburn:InternationalJournalOfEducationalDevelopment:2009}.
In addition to the effect of nation-states and demographics (ethnic groups, rural vs. urban) on urban migrant integration, 
Schiller et al. argue that the role of migrants in the cities depends on the rescaling of the cities themselves~\cite{schiller2009towards}.
Government policy and agenda-setting also play an important role in the integration process \cite{scholten2013agenda}.
Beyond our scope, immigrants (migrants to a new country) and refugees (a subgroup of immigrants) have also received significant interests \cite{becker2011challenge,bean2003america,waters2005assessing,jacobsen2003dual,strang2010refugee}. 
Our work is also related to data-driven studies related to cities, urban computing ~\cite{Quercia:2015:DLW:2736277.2741631,afridi2015social,DBLP:journals/corr/DredzeGRM16,Zheng:2014:UCC:2648782.2629592,jiang2013review,reades2007cellular,zheng2011urban,Hristova:2016:MUS:2872427.2883065}.

\section{Concluding Discussions}
\label{sec:conclusion}

We present the first large-scale study on migrant integration based on telecommunication metadata.
By studying 
the differences between \locals, \migrants, and \newcomers, 
we demonstrate the evolution of a migrant's communication network and geographical locations in the integration process.
Migrants are indeed approaching \locals in most characteristics despite the short time span.

We further formulate prediction problems to distinguish migrants from \locals.
A classifier based on the features that we propose can achieve an F1-score of around 0.82 on \migrants. 
This confirms that migrants are still fairly different from \locals in their behavior patterns, supporting studies on the segregation of migrants.
Meanwhile, we also observe that a larger fraction of \newcomers is classified as \locals over time, partly documenting the integration process.
We hope that our study can encourage more researchers in our community to examine the problem of migrant integration from different perspectives and eventually lead to methodologies and applications that benefit policymaking and millions of migrants.

\small
\vpara{Acknowledgements.} 
We sincerely thank China Telecom for providing the data.
The work is supported by the Fundamental Research Funds for the Central Universities, 973 Program (2015CB352302), NSFC (U1611461, 61625107), and key program of Zhejiang Province (2015C01027). Tan was partly supported by a University of Washington Innovation Award. 
\normalsize

\balance
\bibliographystyle{aaai}
\bibliography{reference}

\end{document}